\documentclass[11pt,onefignum,onetabnum]{siamart220329}

\usepackage{lipsum}
\usepackage{amsfonts}
\usepackage{graphicx}
\usepackage{epstopdf}
\usepackage{algorithm}
\usepackage[utf8]{inputenc}
\usepackage{amsmath}
\usepackage{latexsym}
\usepackage{amssymb}
\usepackage{stmaryrd}
\usepackage{caption}
\usepackage{color}
\usepackage[colorinlistoftodos]{todonotes}
\usepackage{enumerate}
\usepackage{bbm}
\usepackage{bm}
\usepackage{amsopn}
\usepackage[caption=false]{subfig}
\usepackage{comment}
\ifpdf
  \DeclareGraphicsExtensions{.eps,.pdf,.png,.jpg}
\else
  \DeclareGraphicsExtensions{.eps}
\fi

\newsiamremark{remark}{Remark}
\newsiamremark{hypothesis}{Hypothesis}
\crefname{hypothesis}{Hypothesis}{Hypotheses}
\newsiamthm{claim}{Claim}
\newtheorem{assumption}[theorem]{\textit{Assumption}}

\newcommand{\QED}{\Box} 
\newcommand{\rw}{\rightarrow}    
\newcommand{\Real}{\mathbb{R}}

\newcommand{\mB}{{\mathcal B}}
\newcommand{\mP}{{\mathcal P}}

\newcommand{\mS}{{\mathcal S}}

\newcommand{\mF}{{\mathcal F}}
\newcommand{\mE}{{\mathcal E}}

\newcommand{\mN}{{\mathcal N}}

\newcommand{\mR}{{\mathcal R}}
\newcommand{\mO}{{\mathcal O}}
\newcommand{\M}{{\mathcal M}}

\newcommand{\mbP}{\mathbb{P}}   
 
\newcommand{\mbE}{\mathbb{E}}

\newcommand{\sff}{{\sf f}}

\newcommand{\sd}{{\sf d}}

\newcommand{\sg}{{\sf g}}

\newcommand{\X}{\mathtt{X}}
\newcommand{\x}{\mathtt{x}}

\newcommand{\ii}{{(i)}}

\newcommand{\beq}{\begin{equation}}
\newcommand{\eeq}{\end{equation}}
\newcommand{\beqa}{\begin{eqnarray}}
\newcommand{\eeqa}{\end{eqnarray}}
\newcommand{\nn}{\nonumber}

\headers{Sequential discretisation schemes for SDE\MakeLowercase{s}}{\"O. D. Akyildiz, D. Crisan, J. Miguez}

\title{Sequential discretisation schemes for a class of stochastic differential equations and their application to Bayesian filtering\thanks{\funding{This work has been partially supported by the the Office of Naval Research (awards N00014-19-1-2226 and N00014-22-1-2647) and Spain's {\em Agencia Estatal de Investigaci\'on} (ref. PID2021-125159NB-I00 TYCHE).}}}

\author{
\"O. Deniz Akyildiz\thanks{Imperial College London (UK), \email{deniz.akyildiz@imperial.ac.uk}.} 
\and Dan Crisan\thanks{Imperial College London (UK), \email{d.crisan@imperial.ac.uk}.}
\and Joaquin Miguez\thanks{Universidad Carlos III de Madrid (Spain), \email{joaquin.miguez@uc3m.es}.} %and Instituto de Investigaci\'on Sanitaria Gregorio Mara\~n\'on 
}

\begin{document}

\maketitle

\begin{abstract}

We introduce a predictor-corrector discretisation scheme for the numerical integration of a class of stochastic differential equations and prove that it converges with weak order 1.0. The key feature of the new scheme is that it builds up sequentially (and recursively) in the dimension of the state space of the solution, hence making it suitable for approximations of high-dimensional state space models. We show, using the stochastic Lorenz 96 system as a test model, that the proposed method can operate with larger time steps than the standard Euler-Maruyama scheme and, therefore, generate valid approximations with a smaller computational cost. We also introduce the theoretical analysis of the error incurred by the new predictor-corrector scheme when used as a building block for discrete-time Bayesian filters for continuous-time systems. Finally, we assess the performance of several ensemble Kalman filters that incorporate the proposed sequential predictor-corrector Euler scheme and the standard Euler-Maruyama method. The numerical experiments show that the filters employing the new sequential scheme can operate with larger time steps, smaller Monte Carlo ensembles and noisier systems. 

\end{abstract}

%% REQUIRED
%
%% REQUIRED

%%%%%%%%%%%%%%%%%%%%%%%%%%%%%%%%%%%%%%%
%
%%%%%%%%%%%%%%%%%%%%%%%%%%%%%%%%%%%%%%%
\section{Introduction}

%%%
\subsection{Background}

Many systems of interest in physics, engineering or the social sciences are modelled by continuous-time state space models \cite{Hamilton94,Patterson08,Schon11} in a probability space $(\Omega,\mF,\mbP)$. For these models, the dynamics of the $d_x$-dimensional signal of interest (or {\em state}), $X(t)$, are described by a multivariate stochastic differential equation (SDE) \cite{Oksendal07} over a time interval $[0,T]$. Typically, the state cannot be observed directly. Instead, $d_y$-dimensional measurements $Y(t_k)$ are collected {\em instantaneously}, at prescribed time instances $\{ t_k: k=1, 2, \ldots, K \}$. In most practical problems, these observations are noisy, partial and often obtained by a nonlinear transformation of the state $X(t_k)$. Within this framework, the Bayesian filtering problem consists in the computation of the conditional probability law of $X(t)$ given all the available observations up to time $t$, i.e., the data $Y(t_1), Y(t_2), \ldots, Y(t_k)$, with $t_1 < t_2 < \cdots < t_k \le t$.  

This mixed continuous-discrete-time framework can be embedded into a fully discrete-time model where one is interested in computing the conditional probability law of the state at the observation times (see, e.g., \cite{Jazwinski70}). Since any solution of a multivariate SDE is a Markov process, it follows that the sequence $X(t_k)$, $k=0,1,\ldots$, is a Markov chain with transition kernel
\beq
\M_k^*(x_{k-1},\sd x_k):=\mbP(X(t_k)\in \sd x_k | X(t_{k-1})=x_{k-1}).
\nn
\eeq 

The observation process ${Y}$ is typically given by 
\beq
Y(t_k) = h\left( X( t_k ) \right) + U_k
\nn %\label{Ytn}
\eeq
where $\{ U_k: k=1, \ldots, K\}$ is a sequence of independent random variables (r.v.'s) and $h:\Real^{d_x}\mapsto\Real^{d_y}$ is a possibly nonlinear map. Using the language of \cite{Crisan20}, we have a corresponding {\em state space model} $\mS=(\pi_0,\M^*,g)$ which incorporates the probability distribution of the signal at time $t=0$, denoted $\pi_0$, the Markov kernels $\M^*=\{\M_k^*: k=1, \ldots, K\}$ that determine the probabilistic dynamics of the state $X(t_k)$ and the {\em potential} functions $g:=\{g_1,\ldots,g_k\}$ that relate the observation $Y(t_k)$ to the state $X(t_k)$. The potential $g_k(x)$ coincides (up to a proportionality constant) with the probability density function (pdf) of $Y(t_k)$ conditional on $X(t_k)=x$. 

For a fixed sequence of observations $y_1, \ldots, y_K$, the model $\mS$ yields a deterministic sequence of probability measures $\pi_1, \ldots, \pi_K$, where $\pi_k$ describes the probability distribution of $X(t_k)$ conditional on the subsequence $Y(t_1)=y_1, \ldots, Y(t_k)=y_k$. If the sequence $Y(t_1), \ldots, Y(t_K)$ is random, then the state space model $\mS$ generates an associated sequence of random probability measures. Either deterministic or random, the sequence $\{\pi_k\}=\{\pi_1, \ldots, \pi_K\}$ is the solution to the optimal filtering problem \cite{Anderson79}. Hence, the probability measure $\pi_k$ is often referred to as the {\em optimal filter} at time $t_k$.

It has been shown in \cite{Crisan20} that the sequence of optimal filters $\{\pi_k\}$ depends continuously on the components of the state space model $\mS=(\pi_0,\M^*,g)$. This property is crucial. In general, the actual Markov kernels $\M_k^*$ induced by the SDE are intractable and they have to be replaced by {\em approximate} transition kernels $\M_k^h$ that correspond to a numerical discretisation scheme indexed by a time resolution parameter $h>0$ (see \cite{Kloeden95,Kloeden12} for an extensive survey of such schemes). This means that, instead of the original model $\mS=(\pi_0,\M^*,g)$ generating the optimal filters $\{\pi_k\}$, we have an approximate model $\mS^h=(\pi_0,\M^h,g)$, with $\M^h:=\{\M_1^h, \ldots, \M_K^h\}$, generating a sequence of approximate probability laws $\{\pi_k^h\}$. In this paper we analyse the error $\pi_k - \pi_k^h$ when the discretised Markov kernels $\M_k^h$ result from weakly convergent \cite{Kloeden95,Rossler09} numerical schemes, including a new predictor-corrector method.

%%%%%%
%
%%%%%%
\subsection{Discrete-time Bayesian filters}
%The optimal filtering problem can be viewed as the transformation of the {\em prior} distribution $\pi_0$ into the {\em posterior} probability distributions $\pi_k$, conditional on the data $y_1, \ldots, y_k$, by means of Bayes' theorem. Hence, the optimal filter $\pi_k$ is often referred to as a Bayesian filter \cite{Sarkka13}.
   
The numerical computation of the approximate filters $\{\pi_k^h\}$ can be carried out using different methods. Popular families of algorithms include nonlinear Kalman methods \cite{Julier04,Arasaratnam09,Menegaz15}, recursive Monte Carlo methods such as particle filters \cite{Gordon93,Doucet00,Djuric03,Cappe07,Kunsch13,jasra2017multilevel}, ensemble Kalman filters (EnKF's) \cite{Evensen03,Evensen09,Iglesias13,Schillings17} and ensemble variational methods \cite{Zupanski05,Zupanski08,Sakov12,Bannister17,Schillings18,Grudzien22}. These algorithms essentially rely on Bayes' theorem to (numerically) convert $\pi_{k-1}^h$ into $\pi_k^h$ when the observation $y_k$ becomes available and, hence, they are often referred to as Bayesian filters \cite{Sarkka13}.  

Let $\pi_k^{h,M}$ be an approximation of $\pi_k^h$ generated by a (numerical) Bayesian filter of any of the families mentioned above. The superscript $M$ denotes, typically, a parameter that controls the computational cost of the algorithm, e.g., the number of Monte Carlo samples in a particle filter. The analysis of numerical Bayesian filters has focused on a strictly discrete-time setup. In our framework, this means that most authors have investigated the accuracy of $\pi_k^{h,M}$ as an estimator of $\pi_k^h$, rather than the actual posterior law $\pi_k$. One recent exception is \cite{Grudzien20}, which presents a numerical study of the performance of the EnKF combined with several numerical schemes (i.e., several kernels $\M_k^h$), including Euler-Maruyama, Milstein, a 4-stage Runge-Kutta method and a strong order 2.0 Taylor scheme, all of them for the stochastic Lorenz 96 model. Related work can also be found in \cite{Menard21,Pannekoucke20}, where the contribution of model-discretisation errors to the covariance matrix of forecast errors is studied. However, to the best of our knowledge, the question of how the {\em true} conditional laws $\{\pi_k\}$ generated by model $\mS=(\pi_0,\M^*,g)$ are approximated by numerical implementations of Bayesian filters designed for the discrete-time model $\mS^h=(\pi_0,\M^h,g)$ has not been theoretically answered.

%%%
\subsection{Contributions}

We enumerate the three main contributions of the paper:
\paragraph{A novel time-discretisation scheme for SDEs} We introduce a predictor-corrector discretisation scheme\footnote{The new scheme is similar in nature to the predictor-corrector methods in \cite[Section 15.5]{Kloeden95}.} for $d_x$-dimensional diffusion SDEs and prove that it converges with weak order 1.0. The new discretisation scheme differs from classical predictor-corrector methods. Let $X_n$, $n=0, 1, 2, \ldots$, denote the $d_x$-dimensional random sequence generated by the new scheme. Under certain assumptions on the diffusion term of the SDE, $X_n$ is built up by sequentially (and recursively) generating $q$ sub-vectors, each with dimension $m_x=\frac{d_x}{q}$. We use this strategy to produce a first order numerical scheme, for which we provide an explicit convergence proof. However, the same strategy can be used to design similar higher order sequential predictor-corrector schemes. We show numerically, using the stochastic Lorenz 96 system as a test model, that the proposed sequential predictor-corrector Euler method is more robust than the standard (first order) Euler-Maruyama scheme, in the sense that it can operate with larger time steps and, therefore, generate valid sequences $X_n$ with a smaller computational cost.

\paragraph{A convergence analysis for the approximate laws $\{\pi_k^h\}$} In the second part of the paper we turn our attention to the theoretical analysis of the error incurred by replacing the transition kernel $\M_k^*$ of the continuous-time signal $X(t_k)$ by the kernel $\M_k^h$ of the random sequence $X_k$ obtained by running the new sequential predictor-corrector scheme from $t_{k-1}$ to $t_k$. In particular, we analyse the error in the approximation of $\pi_k$ by $\pi_k^h$, prove that $\pi_k^h \rw \pi_k$ as $h\rw 0$ and provide explicit error bounds for the error. Remarkably, our analysis holds not only for the proposed method but also for a broad class of weakly convergent schemes. Hence, it provides theoretical validation for the combination of time-discretisation schemes and discrete-time filters in continuous-time systems. 

\paragraph{A numerical study of the performance of the EnKF in the continuous-discrete-time framework} Finally, we assess the performance of several EnKF's that incorporate the proposed sequential predictor-corrector Euler scheme and the standard Euler-Maruyama method. We carry out a numerical comparison of several algorithms for the stochastic Lorenz 96 system, which is a popular test model in geophysics. For our experiments, we have considered both standard EnKF's that differ only on the numerical scheme used in the prediction step, but also a {\em sequential} EnKF method that takes explicit advantage of the new discretisation scheme. Our computer experiments show that the filters employing the new sequential scheme are numerically more robust, meaning that they can operate with larger time steps, smaller Monte Carlo ensembles, and noisier systems. 

The rest of the paper is organised as follows. The last part of the current section is a summary of the notation used through the manuscript. The sequential predictor-corrector scheme is introduced in Section \ref{sDiscretisationSDEs}. The theoretical analysis of the approximations $\{\pi_k^h\}$ is carried out in Section \ref{sSSM}. Section \ref{sExamples} is devoted to the numerical study of ensemble Kalman filters constructed around different discretisation schemes. Finally, we outline the main conclusions of our work in Section \ref{sConclusions}.

%%%
\subsection{Notation}

We complete this introductory section with a summary of notation used throughout the manuscript.
\begin{itemize}

\item $a \in \Real^d$ denotes a $d$-dimensional column vector with real entries, while $A \in \Real^{d \times m}$ is a real matrix with $d$ rows and $m$ columns. The $r$-th entry of $a$ is denoted $a^r$ and the entry in the $r$-th row and $c$-th column of $A$ is denoted $A^{r,c}$.

\item Let $X:\Omega\mapsto\Real^d$ be a $d$-dimensional random variable (r.v.) defined on a probability space $(\Omega,\mF,\mbP)$. Then $\mbE[X] := \int X \sd \mbP$ denotes the expected value of $X$ and $\| X \|_p = \left( \mbE\left[ \sum_{i=1}^d |X^i|^p \right] \right)^{\frac{1}{p}}$ is the $L_p$ norm of $X$.

\item For $\mR \subseteq \Real^d$, $\mB(\mR)$ denotes the Borel $\sigma$-algebra of open subsets of $\mR$.

\item $\mP(\mR)$ is the class of probability measures on the measurable space $(\mR,\mB(\mR))$.

\item $\delta_{x'}(\sd x)$ is the Dirac delta measure that assigns a unit probability mass to the point $x'$. 

\item Let $\pi$ denote the probability law of the r.v. $Z$ and let $\phi:\Real^d\mapsto \Real$ be a real test function. If $\phi$ is integrable with respect to (w.r.t.) $\pi$, then we denote $\pi(\phi) := \int \phi(z) \pi(\sd z) = \mbE[ \phi(Z) ]$.

\item Let $\alpha = (\alpha_1, \ldots, \alpha_d) \in \left( \mathbb{N} \cup \{0\} \right)^d$ denote a multi-index and let $f : \Real^d \mapsto \Real$ be a real function. If $x = [x_1, \ldots, x_d]^\top$, then $f^{(\alpha)}(x) = \frac{
	\partial^{\alpha_1}
}{
	\partial x_1
} \frac{
	\partial^{\alpha_2}
}{
	\partial x_2
} \cdots \frac{
	\partial^{\alpha_d}
}{
	\partial x_d
} f(x)$ denotes a derivative of order $|\alpha|=\sum_{i=1}^d \alpha_i$.

\item For a constant $B>0$, $C_B^l(\mR)$ denotes the set of real continuous functions $\mR \mapsto \Real$ with derivatives up to order $l$ uniformly bounded by $B$. 

\item We work with a stochastic process $X(t)$ which is $d_x$-dimensional and can be partitioned into $m_x$-dimensional processes denoted $X_0(t), \ldots, X_{q-1}(t)$, where $m_x = \frac{d_x}{q}$.
\item Given a sequence $x_i, x_{i+1}, \ldots, x_{j}$ we use $x_{i:j}$ to denote either the set $\{x_i, x_{i+1}, \ldots, x_j \}$ or the $(j-i+1)$-dimensional vector $[x_i, \ldots, x_j]^\top$.
\end{itemize}

%%%%%%%%%%%%%%%%%%%%%%%%%%%%%%%%%%%%%%%
%
%%%%%%%%%%%%%%%%%%%%%%%%%%%%%%%%%%%%%%%
\section{A sequential predictor-corrector numerical scheme} \label{sDiscretisationSDEs}

%%%%%%
%
%%%%%%
\subsection{The discretisation scheme} \label{ssSTScheme}

Let $W(t)$ denote a $d_x$-dimensional Wiener process defined on a probability space $(\Omega,\mF,\mbP)$, let $T<\infty$ be an arbitrary time horizon and choose two functions $f:\Real^{d_x} \times [0,T]\mapsto\Real^{d_x}$ and $s : \Real^{d_x} \times [0,T]\mapsto\Real^{d_x \times d_x}$. Let $X(t)$, $0 \le t \le T$, be the solution of the It\^o SDE 
\beq
\sd X = f(X,t) \sd t + s(X,t) \sd W.
\label{eqSDE0}
\eeq 
Under certain regularity assumptions (see, e.g., \cite{Oksendal07}), the existence of $X(t)$ can be guaranteed. We assume that the diffusion coefficient $s(X,t) \in \Real^{d_x \times d_x}$ is a block-diagonal matrix,
\beq
s(X,t) = \left[
	\begin{array}{ccc}
	s_0(X,t) &\cdots &0\\
	\vdots &\ddots &\vdots\\
	0 &\cdots &s_{q-1}(X,t)\\
	\end{array}
\right],
\nn
\eeq
where each $s_i(X,t)$, $i \in \{0, ..., q-1\}$, is a $\Real^{d_x} \times[0,T] \mapsto \Real^{m_x \times m_x}$ function and $q = \frac{d_x}{m_x} \ge 1$ is an integer. Similarly, for the $d_x$-dimensional drift coefficient we denote 
$$
f(X,t)=\left[ 
	\begin{array}{c}
	f_0(X,t) \\
	\vdots\\ 
	f_{q-1}(X,t)\\
	\end{array}
\right],
$$ 
where $f_i : \Real^{d_x} \mapsto \Real^{m_x}$. The $r$-th element of the drift function is denoted $f^r$ and the entry in the $r$-th row and $c$-th column of the diffusion coefficient is denoted $s^{r,c}$.

We introduce a predictor-corrector discretisation scheme for Eq. \eqref{eqSDE0} that can be run sequentially and recursively both over time $t$ {\em and} over the coordinates of the system. To describe it explicitly, let us denote
\beq
X(t) = \left[
	\begin{array}{c}
	X_0(t)\\
	\vdots\\
	X_{q-1}(t)\hspace{-2mm}\\
	\end{array}
\right], \
W(t) = \left[
	\begin{array}{c}
	W_0(t)\\
	\vdots\\
	W_{q-1}(t)\hspace{-2mm}\\
	\end{array}
\right],\
V_n = \left[
	\begin{array}{c}
	W_0(nh)-W_0((n-1)h)\\
	\vdots\\
	W_{q-1}(nh)-W_{q-1}((n-1)h)\hspace{-2mm}\\
	\end{array}
\right],
\nn
\eeq 
where $h>0$ is a time step parameter, and $X_i(t)$ and $W_i(t)$ are $m_x$-dimensional vectors (recall that $m_xq=d_x$). The vector $V_n$ is Gaussian-distributed, with mean $\mbE[V_n]=0$ and covariance matrix $\mbE[ V_nV_n^\top ] = hI_{d_x}$, where $I_{d_x}$ is the $d_x \times d_x$ identity matrix. This is denoted as $V_n \sim \mN(0,hI_{d_x})$. We may also decompose $V_n$ into a collection of $m_x$-dimensional r.v.'s $V_{i,n} = W_i(nh)-W_i((n-1)h) \sim \mN(0,h I_{m_x})$, for $i=0, \ldots, q-1$.

We construct a discrete-time approximation of the process $X(t)$, $0\le t \le T$, over the grid $\left\{ t_n = nh: h = \frac{T}{N} \text{ and } n=0, 1, \ldots, N \right\}$, using the predictor-corrector scheme outlined in Algorithm \ref{algSTD}. It yields a sequence 
$
X_n = \left[
	\begin{array}{c}
	X_{0,n}\\ 
	\vdots\\
	X_{q-1,n}\\
	\end{array}
\right]
$, where $X_n$ is an estimate of $X(t_n)$ and, correspondingly, $X_{i,n}$ is an estimate of $X_i(t_n)$. The procedure consists of a deterministic prediction step, which yields an auxiliary estimate $\hat X_n$ at time $t_n$, followed by a corrector step in which the estimates $X_{i,n}$ are computed sequentially, and recursively, for $i=0, \ldots, q-1$. 

An intuitive way to obtain Algorithm \ref{algSTD} is to start with the backward Euler scheme \cite{Kloeden95}
\beq
X_n = X_{n-1} + hf_n(X_n) + s_{n-1}(X_{n-1}) V_n.
\nn
\eeq
This is an implicit method that requires to solve $d_x$ nonlinear equations at each time step. Instead of taking this approach, we first convert it into a predictor-corrector scheme,
\beq
X_n = X_{n-1} + hf_n(\hat X_n) + s_{n-1}(X_{n-1})V_n,
\nn
\eeq
where $\hat X_n = X_{n-1} + hf_{n-1}(X_{n-1})$. Finally, instead of computing the corrected states $X_n$ in a single shot, we perform the corrections one $m_x$-dimensional block at a time, 
\beqa
X_{0,n} &=& X_{0,n-1} + hf_{0,n}(\hat X_{0:q-1,n}) + s_{0,n-1}(X_{n-1})V_{0,n},\nn\\
X_{1,n} &=& X_{1,n-1} + hf_{1,n}(X_{0,n},\hat X_{1:q-1,n}) + s_{1,n-1}(X_{n-1})V_{1,n},\nn\\
%X_{2,n} &=& X_{2,n-1} + hf_n(X_{0:1,n},\hat X_{2:q-1,n}) + s_{1,n-1}(X_{n-1}),\nn\\
&\vdots&\nn\\
X_{q-1,n} &=& X_{2,n-1} + hf_{q-1,n}(X_{0:q-2,n},\hat X_{q-1,n}) + s_{1,n-1}(X_{n-1})V_{q-1,n},\nn
\eeqa
using the corrected states up to $i-1$, $X_{0:i-1,n}$, in order to compute $X_{i,n}$ and denoting $f_{i,n}(\cdot) := f_i(\cdot,t_n)$, $s_{i,n}(\cdot):=s_i(\cdot,t_n)$. Because of the sequential correction procedure above, we refer to this method as a sequential predictor-corrector scheme. For conciseness, and because it is built around the backward Euler scheme, we use the term `sequential Euler' for Algorithm \ref{ssSTScheme} hereafter. (Note that Algorithm \ref{algSTD} is still valid for $q=1$, although in this case the procedure is not `sequential' any more because all state variables are updated together). 

%%%
\begin{remark}
Although in this paper we restrict our attention to the sequential Euler scheme, the methodology outlined in this section can be applied in a straightforward way to other implicit schemes in order to obtain sequential predictor-corrector methods of different types.
\end{remark}

%%%
\begin{remark}
The predictor step a) in Algorithm \ref{algSTD} is deterministic. A stochastic predictor can also be (easily) implemented. However, we have not been able to show any clear gain with such modification and, therefore, we have opted for the deterministic predictor which makes both the implementation and the analysis simpler.
\end{remark}

%%%
\begin{algorithm}[htb]
\caption{Sequential predictor-corrector Euler scheme for $d_x$-dimensional SDEs.}
\label{algSTD}
\begin{enumerate}
\item \textbf{Initialisation}: let $X_0 = X(0)$ and denote $X_0 =[X_{0,0},\dots,X_{q-1,0}]^T$.
%$
%X_0 = \left[ 
%	\begin{array}{c}
%$	X_{0,0}\\
%	\vdots\\
%	X_{q-1,0}\\
%	\end{array}
%\right]$.
\item \textbf{Sequential step:} for $n = 1, \ldots, N$:
	\begin{enumerate}[a)]
	\item {\em Predictor}: compute an auxiliary estimate $\hat X_n \in \Real^{d_x}$ via the Euler step
	\beq
	\hat X_n = X_{n-1} + hf_{n-1}(X_{n-1}),
	\label{eqAuxX}
	\eeq 
	where $f_n(\cdot) := f(\cdot,t_n)$.
	\item {\em Corrector}: For $i = 0, \ldots, q-1$, compute
	\beqa
	X_n[i] &=& \left[
		\begin{array}{c}
		X_{0:i-1,n}\\
		\hat X_{i:q-1,n}\\
		\end{array}
	\right], \nn\\
	X_{i,n} &=& X_{i,n-1} + hf_{i,n}(X_n[i]) + s_{i,n-1}(X_{n-1}) V_{i,n},
	\nn
	\eeqa
	where $X_n[0] = \hat X_n$ by convention, $s_{i,n}(\cdot):=s_i(\cdot,t_n)$ and $f_{i,n}(\cdot) := f_i(\cdot,t_n)$.% and $X_{m:r,n} = \{ X_{m,n}, \ldots, X_{r,n}\}$, with $X_{m:r,n}=\emptyset$ if $m<r$.
	\end{enumerate}
\end{enumerate}
\end{algorithm}
%%%

%%%%%%
%
%%%%%%
\subsection{Weak convergence} \label{ssWeak}

We are interested in the approximation of the random variables $X(t_n)$ in terms of their marginal probability laws, denoted by $\mu_n$. In this section we prove that the sequential Euler scheme, with time step $h=\frac{T}{N}>0$, yields a sequence $X_n$ such that $\lim_{h\rw 0} \mu_n^h = \mu_n$ in a suitable quantitative sense, where $\mu_n^h$ is the law of $X_n$. For this, we adapt a weak convergence criterion of \cite{Kloeden95} to suit the new scheme. In particular, we use stochastic Taylor expansions in order to derive appropriate discrete-time approximations. As with strong approximations, the desired order of convergence also determines the truncation that must be used. However, this is different from the truncation required for the strong convergence of the same order, in general involving fewer terms (see, e.g., \cite{Kloeden95} for details).

Theorem \ref{thWeak} below guarantees that the sequential Euler scheme converges with weak order 1 when $h \rw 0$. Our analysis relies on the following assumption.

%%%
\begin{assumption} \label{asW}
The coefficients of the SDE \eqref{eqSDE0} are uniformly bounded, i.e., 
\beq
\sup_{i,t,x} |f^i(x,t)| \vee |s^{i,j}(x,t)| < \infty,
\nn
\eeq 
where $a \vee b$ denotes the maximum between $a$ and $b$. Moreover $f^i(\cdot,t), s^{i,j}(\cdot,t) \in C_B^4\left( \Real^{d_x} \right)$ for some fixed (but arbitrary) constant $B<\infty$, every $t \in [0,T]$ and every $i,j \in \{0,...,d_x-1\}$. 
\end{assumption}
%%%

%%%
\begin{theorem} \label{thWeak}
If Assumption \ref{asW} holds, then, for any test function $\phi \in C_B^4\left( \Real^{d_x} \right)$,
\beq
\left| 
	\mbE\left[ 
		\phi\left( X(T) \right) 
	\right] - \mbE\left[ \phi\left(
			X_N 
		\right) 
	\right] 
\right| \leq C h  
\label{eqWeakOrder}
\nn
\eeq
where the constant $C=\mO(Td_x^2)<\infty$ is independent of the time step $h=\frac{T}{N}$ and the initial value $X(0)=X_0$. 
\end{theorem}
%%%

\begin{proof} 
See Appendix \ref{apProofThWeak}.
\end{proof}

%%%
\begin{remark} \label{rmMu}
At time $t_n=nh$ we have
$
\mbE\left[ 
	\phi\left( X_n \right) 
\right] = \mu_n^h(\phi)
$ and, similarly, 
$
\mbE\left[ 
	\phi\left( X(t_n) \right) 
\right] = \mu_n(\phi)
$. Theorem \ref{thWeak} guarantees that $\sup_{n \le N} \left| \mu_n^h(\phi) - \mu_n(\phi) \right| = \mathcal{O}(h)$ for any test function $\phi \in C_B^4(\Real^{d_x})$. 
\end{remark}
%%%

%%%
\begin{remark}
If we relax Assumption \ref{asW} to request only that $f^i(\cdot,t), s^{i,j}(\cdot,t) \in C_B^{4}\left( \Real^{d_x} \right)$ for all $t \in [0,T]$ and $i,j \in \{0,...,d_x-1\}$ (i.e., the coefficients themselves are no longer assumed to be uniformly bounded), then we have the inequality
\beq
\left| 
	\mbE\left[ 
		\phi\left( X(T) \right) 
	\right] - \mbE\left[ \phi\left(
			X_N 
		\right) 
	\right] 
\right| \leq C_{x,0} h,  
\label{eqWeakOrder_2}
\nn
\eeq
where the constant $C_{x,0}<\infty$ may depend on the initial condition $X(0)$ (but not on the time step $h$). 
\end{remark}

%%%
\begin{remark}
If the drift and diffusion functions, $f(\cdot,t)$ and $s(\cdot,t)$, respectively, are not sufficiently smooth then the scheme can still be shown to converge, albeit with a lower order. This can be done in a similar manner as in \cite{Kloeden95}. 
%Yes this is correct: \ccyan{This ref. is Kloeden's book. Is this ok?}
\end{remark}

%%%%%%
%
%%%%%%
% \subsection{General sequential predictor-corrector schemes}
% \label{ssGeneralSPC}
% The method described in Section \ref{ssSTScheme} above is a PC scheme based on the backward Euler method, with the peculiarity that the correction step is sequential and recursive. The same strategy can be applied to obtain a sequential PC method from virtually any implicit scheme. 

% Let us take, for example, the implicit ...

% For simplicity, in the rest of the paper we focus on the sequential Euler scheme outlined in Section \ref{ssSTScheme}. 

%%%%%%
%
%%%%%%
\subsection{Example: The stochastic Lorenz 96 model} \label{sSL96}
%\label{ssSL96}
%\label{sssNumResSchemes}

In order to illustrate the application of the proposed methodology we use it for the time discretisation of a stochastic Lorenz 96 model with multiplicative noise (see e.g. \cite{Grudzien20}). To be specific, we work with the $d_x$-dimensional SDE
\beq
\sd X^i = -X^{i-1} X^{i-2}  - X^{i+1} - X^i + F + \sigma X^i \sd W^i, \quad i=0, \ldots, d_x-1,
\label{eqL96}
\eeq
where $F$ is a forcing constant, $W^i$, $i=0, \ldots, d_x-1$, are standard Wiener processes and $\sigma$ is a constant diffusion factor. The operations on the index $i$ are performed modulo $d_x$, hence, for $0 \le k < d_x$ we have $X_{-k}=X_{d_x-k}$ and $X_{d_x-1+k}=X_k$. Choosing $F>6$ makes the dynamics of the {\em deterministic} Lorenz 96 model (obtained with $\sigma=0$) chaotic and, therefore, very sensitive to small perturbations and hard to predict. We set $F=8$ for the computer experiments in this paper.

The SDE is discretised using the sequential Euler scheme in Section \ref{ssSTScheme}, with time step $h>0$. The partition of the state vector $X_n$ is carried out in 1-dimensional components, i.e., $m_x = 1$ and $q=d_x$ when comparing with the general case of Section \ref{sDiscretisationSDEs}.
We evaluate, numerically, the performance of the proposed sequential Euler scheme and the standard (explicit) Euler-Maruyama method as we vary the time step $h>0$ and the diffusion factor $\sigma>0$.

\begin{figure}[htb]
\centerline{
    \subfloat[$h=10^{-3}$]{
        \label{fTraj_a}
        \includegraphics[width=0.4\linewidth]{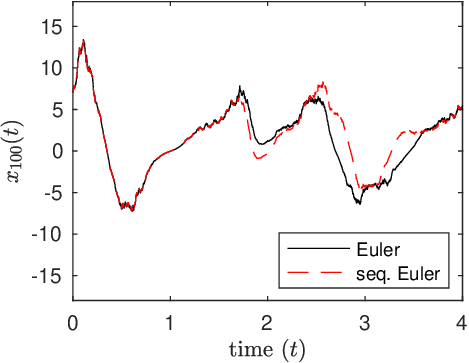}
    }
    \subfloat[$h=5 \times 10^{-3}$]{
        \label{fTraj_b}
        \includegraphics[width=0.4\linewidth]{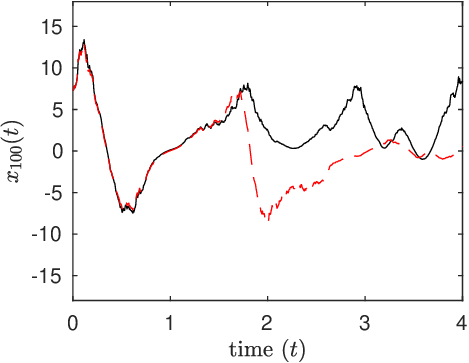}
    }
}
\centerline{
    \subfloat[$h=10^{-2}$]{
        \label{fTraj_c}
        \includegraphics[width=0.4\linewidth]{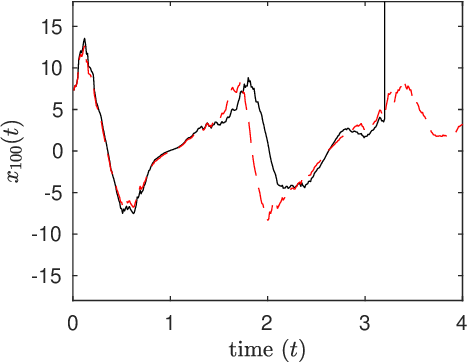}
    }
    \subfloat[$h=5 \times 10^{-2}$]{
        \label{fTraj_d}
        \includegraphics[width=0.4\linewidth]{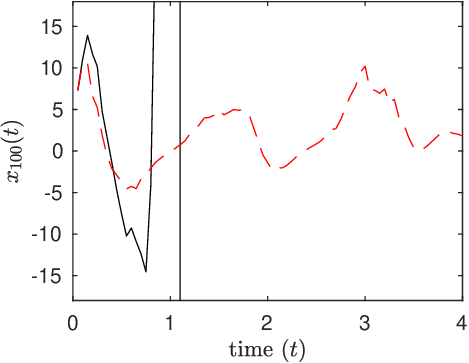}
    }
}
\caption{Sample trajectories with the standard Euler and sequential Euler methods. The variable plotted is $x_{100}(t)$ and the overall dimension is $d_x=200$. The diffusion factor is $\sigma=\sqrt{1/2}$. The horizontal axis is continuous time, $t \in [0,4]$. The time step, $h$, is indicated in the caption of each plot. For $h=10^{-2}$ and $h=5 \times 10^{-2}$ the standard Euler scheme does not complete the simulation. }
\label{fTraj}
\end{figure}

In Figure \ref{fTraj} we plot simulated trajectories for the $100$-th entry, $X_{100}(t)$, of the stochastic Lorenz 96 model \eqref{eqL96} with dimension $d_x=200$. These trajectories are generated using Algorithm \ref{algSTD} (labeled `seq. Euler') and the standard Euler-Maruyama method (labeled `Euler') with a common initial condition, diffusion factor $\sigma=\sqrt{1/2}$ and increasing values of the time step $h>0$. The length of the simulation is $T=4$ continuous-time units and the realisation of the Wiener process $W(t)$, $t\in [0,T]$ is the same for the three algorithms and all values $h$. 

We observe that the simulated trajectories are very similar (nearly identical) at the beginning of the simulation interval and then they progressively depart. When $h$ is sufficiently small (e.g., Figure \ref{fTraj_a}) the trajectories stay close for a longer time, while when we choose a larger value of $h$ (e.g., Figure \ref{fTraj_b}) the algorithms yield trajectories that separate clearly after $\approx 2.0$ continuous-time units. For larger time steps, $h=0.01$ in Figure \ref{fTraj_c} and $h=0.05$ in Figure \ref{fTraj_d}, the simulations with the standard Euler method cannot be completed because the trajectories ``explode'', i.e., they increase quickly in absolute value until they overflow the machine representation capability. Remarkably, the sequential Euler scheme still yields complete simulated trajectories in Figs. \ref{fTraj_c} and \ref{fTraj_d}. This experiment suggests that the sequential Euler scheme is numerically more robust than the standard Euler method as it can run over a coarser time grid.  \\[-5mm]

We have carried out a more detailed computer experiment to assess the numerical robustness of the two methods to variations in the time step $h$ and the diffusion factor $\sigma$. In particular, we have run simulations of length $T=2.0$ continuous-time units, with independent random initial conditions, independent realisations of the Wiener processes (for each scheme) and dimension of the state $d_x=200$. 

For each simulation, we run a standard Euler scheme with time step $h_o=10^{-6}$ and then the standard and sequential Euler schemes with time steps ranging from $h=10^{-4}$ to $h=10^{-1}$. The reference Euler method with step $h_o=10^{-6}$ yields a sequence $X_n$ that we use as a proxy for the ground truth signal $X(nh_o)$. We evaluate the performance of the schemes in terms of weak errors $\left| \mbE[\phi(X(T))] - \mbE[\phi(X_N)] \right|$, where we choose the test function $\phi$ to be the Euclidean norm, $\phi(X)=\|X\|^2=\sqrt{\sum_i \left(X^i\right)^2}$. The quantity $\mbE\left[\| X(T) \|^2\right]$ is approximated by averaging $J$ independent runs of the reference Euler scheme, i.e.,
\beq
\mbE[\left\| X(T) \|^2 \right] \approx \frac{1}{J}\sum_{j=1}^J \| X_{\frac{T}{h_o}}^{(j)} \|^2 =: \ell^2(T,h_o), 
\nn
\eeq
where $X_{\frac{T}{h_o}}^{(j)}$ is the last element of the sequence in the $j$-th simulation. Similarly, with a given scheme (either Euler or sequential Euler) and time step $h$, $\mbE\left[ \| X_N \|^2 \right]$ is estimated as
\beq
\mbE[\left\| X_N \|^2 \right] \approx \frac{1}{J}\sum_{j=1}^J \| X_{\frac{T}{h}}^{(j)} \|^2 =: \ell^2(T,h), 
\nn
\eeq
where $X_{\frac{T}{h}}^{(j)}$ is the last element of the sequence $X_n$. The normalised weak error is approximated by the quantity $\mE(T,h)=\left| \ell^2(T,h_o)-\ell^2(T,h) \right| / \ell^2(T,h_o)$. We have run $J=10,000$ independent simulations for each scheme and each value of $h$. 

In order to choose the initial condition for each simulation, we first run a standard Euler scheme for the deterministic Lorenz 96 model in the interval $t \in [0,10]$, with time step $h=10^{-4}$. Then, for each $j=1, \ldots, J$, we choose one point at random, with uniform probabilities, in the generated trajectory and use it as the initial condition $X_0^{(j)}$ for the Euler and sequential Euler methods. In this way we ensure that the simulations are started close to the model attractor and avoid transient phenomena. 

\begin{figure}[htb]
\subfloat[$\sigma=\sqrt{1/4}$]{
         \label{fSchemes_a}
         \includegraphics[width=0.315\linewidth]{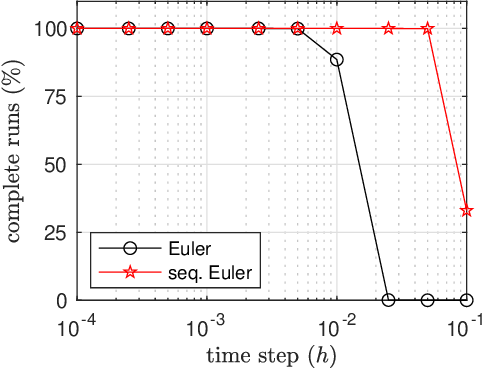}
         \includegraphics[width=0.315\linewidth]{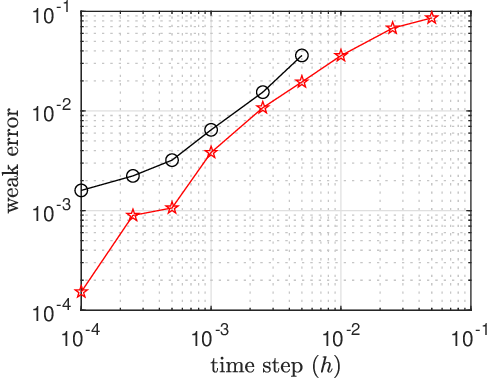}
         \includegraphics[width=0.315\linewidth]{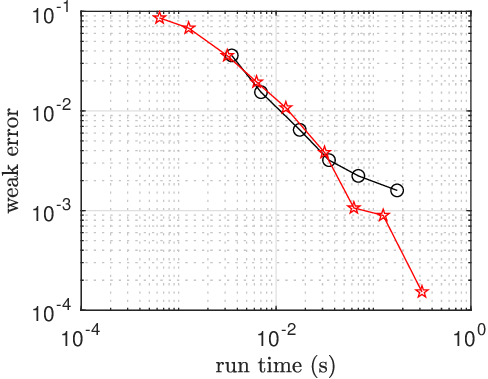}
}\\
\subfloat[$\sigma=1$]{
         \label{fSchemes_b}
         \includegraphics[width=0.315\linewidth]{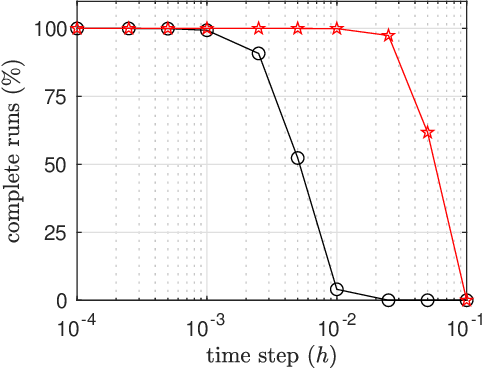}
         \includegraphics[width=0.315\linewidth]{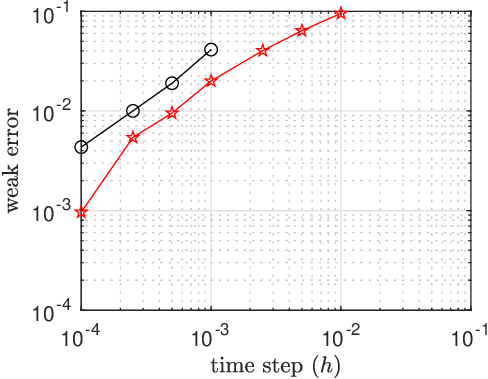}
         \includegraphics[width=0.315\linewidth]{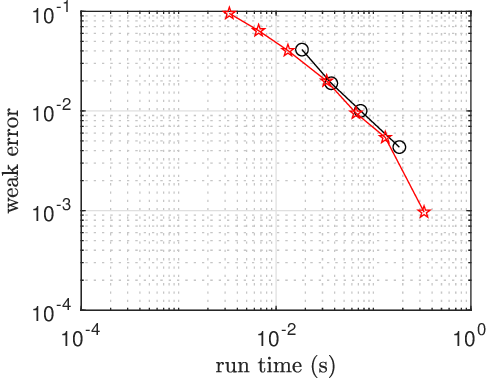}
     }\\
    \caption{Performance of the standard Euler-Maruyama (`Euler') and the sequential predictor-corrector Euler (`seq. Euler') schemes. All graphs are averaged over 10,000 independent simulation runs. The length of the simulation interval is $T=2$ and the dimension of $X(t)$ is $d_x=200$. \textbf{Left}: Percentage of complete simulations. \textbf{Middle}: Normalised weak error versus the time step $h$. \textbf{Right}: Normalised weak error versus run-time in seconds.}
    \label{fSchemes}
\end{figure}

Figure \ref{fSchemes} displays the results of the computer experiment when the diffusion factor is either $\sigma=\sqrt{1/4}$ (Figure \ref{fSchemes_a}) or $\sigma=1$ (Figure \ref{fSchemes_b}). In each panel, the figure on the left shows the percentage of complete runs versus time step $h$ for the Euler (black) and sequential Euler (red) methods. We see that the latter scheme can run with a time step one order of magnitude larger than the standard Euler method. The weak errors $\mE(T,h)$ are also smaller for the sequential Euler scheme for each value of $h$ and $\sigma$, as shown by the plots in the middle. When the weak errors are plot versus the average run times of the algorithms (plots on the right) we observe that the performance is similar but the sequential Euler scheme can operate over a larger range of run times. In particular, it can deliver valid sequences (where $\mE(T,h)<10^{-1}$) with an average run time significantly smaller than the standard Euler method.  

%%%%%%%%%%%%%%%%%%%%%%%%%%%%%%%%%%%%%%%
%
%%%%%%%%%%%%%%%%%%%%%%%%%%%%%%%%%%%%%%%
\section{State-space models} \label{sSSM}

%%%%%%
%
%%%%%%

\subsection{Observations over a time grid} \label{ssOOTG}

Assume that the signal $X(t)$ can be partially observed at time instants $t_k'$, $k=1, \ldots, K$, such that $ 0 < t_1' < \ldots < t_K' = T$. To be specific, the observations are $d_y$-dimensional r.v.'s of the form
\beq
Y_k = b(X(t_k'),t_k') + U_k, \quad k=1, \ldots, K,
\nn %\label{eqObservation0}
\eeq 
where $b:\Real^{d_x}\times [0,T] \mapsto \Real^{d_y}$ is an observation function and $U_k$ is a sequence of independent, zero-mean, $d_y$-dimensional real r.v.'s. If we assume that $U_k$ has a pdf $\bar g_k:\Real^{d_y}\mapsto (0,\infty)$ then we can construct the potential function of the state $X(t_k')=x$  for a given observations $Y_k=y$ as 
\beq
g_k(x) \propto \bar g_k( y - b_k(x) ),
\label{eqLikelihoods}
\eeq
where $b_k(x) := b(x, t_k')$. We also assume, without loss of generality, that $\sup_{x,k} |g_k(x)| \le 1$ (see \cite{DelMoral01c}).

%%%
\begin{remark}
For notational simplicity, we assume that the grid of observation times $\{t_k': k=1, \ldots, K \}$ can be aligned with the discretisation grid $\{ t_n = nh: n=0, \ldots, N \}$ of the numerical scheme, i.e., there are integers $0 < n_1 < \ldots < n_K = N$ such that $t_k' = n_kh$. The size $K$ of the observation grid is independent of the choice of the time step $h$ and, typically, $K<<N$.
\end{remark}

%%%%%%
%
%%%%%%
\subsection{Continuous- and discrete-time Markov kernels} \label{ssKernels} 

Both the continuous-time It\^o process $X(t)$ and its discrete-time approximation $X_n$ are Markov, i.e., for any Borel set $A \subset \Real^{d_x}$, any time $t \in (0,T]$ and any function $x:[0,T] \mapsto \Real^{d_x}$, we have
\beq
\mbP\left( X(t) \in A | X(\tau) = x(\tau), \text{ for } \tau \in [0, t-h] \right) = 
\mbP\left( X(t) \in A | X(t-h) = x(t-h) \right)
\nn
\eeq
for the process $X(t)$, while for the random sequence $X_n$
\beq
\mbP\left( X_n \in A | X_l = x(lh), \text{ for } l = 0, \ldots, m \right) = 
\mbP\left( X_n \in A | X_m = x(mh) \right)
\nn
\eeq
for any $m<n$. Therefore, the dynamics of the process $X(t)$ over the observation grid $t_k' = n_kh$, $k=1, \ldots, K$, can be described by the Markov kernels
\beq
%\M_1^*(X_0,A) &:=& \mbP\left( X(t_1') \in A | X(0) = X_0 \right), \quad \text{and}\nn\\
\M_k^*(x,A) := \mbP\left( X(t_k') \in A | X(t_{k-1}') = x \right), \quad \text{for} \quad k=1, \ldots, K,
\label{eqK*} 
\eeq
where $t_0'=0$. We shall prove that, under the assumptions of Theorem \ref{thWeak}, these exact kernels can be approximated by
\beq
\M_k^h(x,A) := \mbP\left( \X_k \in A | \X_{k-1} = x \right), \quad \text{for} \quad \quad k=1, \ldots, K, \label{eqKh}
\eeq
which generate the subsequence $\X_0= X_0, \X_1=X_{n_1}, \ldots, \X_K=X_{n_K}$.

Let $\mu$ and $\pi$ be two probability measures on $\mP(\Real^d)$. If we denote
\beq
D(\mu,\pi) := \sup_{\phi \in C_B^4(\Real^d) : \|\phi\|_\infty \le 1} | \mu(\phi)-\pi(\phi) |,
\label{defDk}
\eeq
then it can be shown that $D(\cdot,\cdot)$ is a proper metric on the space of probability measures $\mP(\Real^d)$. Moreover, we can state the lemma below, which is a straightforward consequence of Theorem \ref{thWeak}.
%%%
\begin{lemma} \label{lmConvK}
If Assumption \ref{asW} holds, then there is a finite constant $C<\infty$, independent of $h = \frac{T}{N}$ and $x$, such that 
\beq
\sup_x D\left( 
	\M_k^*(x,\cdot), \M_k^h(x,\cdot) 
\right) \le C h
\label{eqConvK}
\eeq
for every $k = 1, \ldots, K$. In particular, $\lim_{h \rw 0} D\left( \M_k^*(x,\cdot), \M_k^h(x,\cdot) \right) = 0$ uniformly on $x \in \Real^{d_x}$ and $k = 1, \ldots, K$. 
\end{lemma}

%\cred{Does the constant $C$ depend on $x$? In principle, it does. We have to clarify Theorem \ref{thWeak} to see this more explicitly.}

%%%
\begin{remark}
We write $\lim_{h\rw 0} \M^h_x = \M^*_x$ to denote that 
\beq
\lim_{h\rw 0} D(\M_k^h(x,\cdot),\M_k^*(x,\cdot)) = 0
\nn
\eeq 
for every $k=1, \ldots, K$. Intuitively, the kernels $\M_k^h$ (that generate the random sequence $\X_k=X_{n_k}$) converge to the kernels $\M_k^*$ that generate the true signal process $X(t)$ on the time grid $t_k'=n_kh$. This argument holds for any weakly-convergent numerical scheme, not just the sequential Euler scheme in Algorithm \ref{algSTD}. 
\end{remark}
%%%

%%%%%%
%
%%%%%%
\subsection{Exact and approximate state-space Markov models}

We refer to the discrete-time random dynamical system described by
\begin{itemize}
\item the initial condition $\X_0^*=X(0)$, with {\em a priori} law $\pi_0$,
\item the Markov sequence $\X_k^*$ generated by the kernels $\M_k^*$, $k=1, \ldots, K$,
\item the potential functions $g_k$, $k=1, \ldots, K$, described by Eq. \eqref{eqLikelihoods},
\end{itemize}
as the {\em exact} state space model (SSM) for the continuous-time signal $X(t)$ with arbitrary (but fixed) observations $\{ Y_k=y_k: k=1, \ldots, K\}$. We denote it by the triple $\mS=(\pi_0, \M^*, \sg)$, where $\M^* := \{ \M_1^*, \ldots, \M_K^* \}$ and $\sg := \{ g_1, \ldots, g_K \}$. 

The term `exact' points out that the {\em a priori} law of $\X_k^*$ is
\beq 
\eta_k(\sd \x_k) := \int \cdots \int 
\M_k^*(\x_{k-1},\sd \x_k) 
\prod_{i=1}^{k-1} \M_i^*(\x_{i-1}, \sd \x_i ) 
\pi_0(\sd \x_0),
\nn
\eeq
which coincides with the {\em a priori} law $\mu_k$ of $X(t_k')$ by the definition of the kernel $\M_k^*$ in Eq. \eqref{eqK*}. Therefore, $\eta_k = \mu_k$ and, for any integrable test function $\phi : \Real^{d_x} \mapsto \Real$,
\beq
\mbE\left[ 
	\phi\left(
		X(t_k')
	\right) 
\right] = \mu_k(\phi) 
= \eta_k(\phi)
= \mbE\left[
	\phi(\X_k^*)
\right]. \nn
\eeq

We can construct an {\em approximate} SSM for the discrete-time sequence $\X_k$ generated by the sequential Euler scheme. In particular, we choose 
\begin{itemize}
\item the same initial condition $\X_0=X(0)$ with law $\pi_0$,
\item the Markov sequence $\X_k$ generated by the kernels $\M_k^h$, and
\item the same likelihoods $g_k$, $k=1, \ldots, K$, as in the exact SSM.
\end{itemize}
We describe this model by the triple $\mS^h=(\pi_0,\M^h,\sg)$, where $\M^h :=\{ \M_1^h, \ldots, \M_K^h\}$. By construction, the {\em a priori} law of $\X_k$ is $\mu_k^h$ and, from Remark \ref{rmMu}, $\lim_{h \rw 0} \mu_k^h = \mu_k$, i.e., the prior law of $\X_k$ converges to the prior law of $\X_k^*=X(t_k')$.

%%%%%%
%
%%%%%%
\subsection{Marginal posterior laws}

Since $\X_0^*$ has marginal law $\pi_0$, the one-step-ahead predictive probability law of $\X_1^*$ can be easily obtained from $\pi_0$. In particular, 
\beq
\xi_1(\sd x_1) = \mbP( \X_1^* \in \sd x_1) = \int \M_1^*(x_0,\sd x_1) \pi_0(\sd x_0)
\nn
\eeq
and we denote $\xi_1 = \M_1^*\pi_0$ for conciseness. Then, given an observation $Y_1=y_1$ and the resulting likelihood function $g_k$, Bayes' theorem yields the posterior marginal law 
\beq
\pi_1(\sd x_1) := \mbP( X_1^* \in \sd x_1 | Y_1=y_1 ) = \frac{
	g_1(x_1) \xi_1(\sd x_1)
}{
	\xi_1(g_1)
}
\nn
\eeq
and we denote $\pi_1 = g_1 \cdot \xi_1 = g_1 \cdot \M_1^*\pi_0$ for conciseness. By induction, one can construct the posterior laws $\xi_k$ and $\pi_k$ recursively, as
\beq 
\xi_k = \M_n^* \pi_{k-1} 
\quad \text{and} \quad
\pi_k = g_k \cdot \xi_k,
\\
\eeq
for each $k = 1, \ldots, K$. See, e.g., \cite{Crisan20} for additional details. The law $\pi_k$ is a posterior probability distribution of $\X_k^*$ conditional on the observations $Y_{1:k} = y_{1:k}$. We refer to $\pi_k$ as the optimal filter at discrete time $k$. 
%%%
%\begin{remark}
%Since the prior probability laws $\eta_1$ and $\mu_1$ of $X_1^*$ and $X(t_1')$, respectively, coincide (indeed, $\eta_1=\mu_1=\xi_1$), then it is apparent that $\pi_1$ is also the posterior law of $X(t_1')$ conditional on  the observation $Y_1=y_1$. 
%A simple induction argument shows that the laws $\xi_k$ and $\pi_k$ also characterise the probability distribution of $X(t_k')$ conditional on $Y_{1:k-1}=y_{1:k-1}$ and $Y_{1:k}=y_{1:k}$, respectively. Hence, $\pi_k$ is also the optimal filter at continuous time $t_k$. 
%\end{remark}
%%%

By the same argument as for the exact model $\mS$ one can construct the sequence of posterior laws $\xi_k^h$ and $\pi_k^h$ for the SSM $\mS^h=(\pi_0,\M^h,\sg)$. Indeed, one obtains
\beq
\xi_k^h = \M_n^h\pi_{k-1} 
\quad \text{and} \quad
\pi_k^h = g_k \cdot \xi_k^h = g_k \cdot \M_k^h\pi_{k-1}^h
\quad
\text{
for $k=1, \ldots, K$
}.
\label{eqXIPIh}
\eeq

Following an argument similar to \cite[Lemma 2.4]{Crisan20} it is possible to prove that $\xi_k^h \stackrel{h\rw\infty}{\longrightarrow} \xi_k$ and $\pi_k^h \stackrel{h\rw\infty}{\longrightarrow} \pi_k$ under suitable regularity assumptions. This is made precise by Theorem \ref{thMarginals} below.

%%%
\begin{theorem} \label{thMarginals}
Let $Y_{1:K} = y_{1:K}$ be an arbitrary but fixed sequence of observations, let Assumption \ref{asW} hold and choose a test function $\phi\in C_B^4(\Real^{d_x})$. If, for every $k=1, \ldots, K$,
\begin{itemize}
\item[(i)] $g_k>0$, $\| g_k \|_\infty \le 1$ and $g_k \in C_B^4(\Real^{d_x})$, and %for every $k=1, \ldots, K$, and
\item[(ii)] $\bar\phi_k \in C_B^4(\Real^{d_x})$, where
$
\bar \phi_k(x_{k-1}) = \int \phi(x_k) \M_k^*(x_{k-1},\sd x_k),
$
\end{itemize}
then there are finite constants $\{ \bar C_k, C_k \}_{k=1}^K$ such that
\beq
D( \xi_k, \xi_k^h) \le \bar C_k h \quad \text{and} \quad D( \pi_k, \pi_k^h ) \le C_k h \quad \text{for $k=1, \ldots, K$}.
\eeq
\end{theorem}
%%%

\begin{proof}
See Appendix \ref{apProofThMarginals}.
\end{proof}

%%%

\begin{remark}
Condition $(ii)$ in the statement of Theorem \ref{thMarginals} is quite natural. Observe that  $\|\bar\phi_k\|_\infty \le \|\phi\|_\infty$. Moreover, $\bar\phi_k$ has the representation 
\beq
\bar\phi_k={\mathbb E}[\phi (X_{t_k'}(t_{k-1},x_{k-1})] 
\nn
\eeq
where $t \mapsto X_{t}(t_{k-1},x_{k-1})$ is the solution at time $t$ of the SDE \eqref{eqSDE0} that starts from $x_{k-1}$ at time $t_{k-1}'$. By differentiating with respect to $x_{k-1}$ under the expectation one can deduce a probabilistic representation for $\bar \phi_k^{(\alpha)}$ of the form 
\beq
\bar \phi_k^{(\alpha)}=\sum_{|\beta|\le |\alpha|}
{\mathbb E}\left[\phi^{(\beta)}\left(X_{t_k'}(t_{k-1}',x_{k-1})\right)
K^{\alpha,\beta}(X)\right] 
\nn
\eeq
where $K^{\alpha,\beta}(X)$ are sums of products of partial derivatives of the stochastic flow $t\mapsto X_{t}(t_{k-1}',x_{k-1})$ with respect to $x_{k-1}$. It follows that 
\beq
\|\bar \phi_k^{(\alpha)} \|_\infty=\sum_{|\beta|\le |\alpha|}\|\phi ^{(\beta)} \|_\infty
{\mathbb E}[|K^{\alpha,\beta}(X)|]. 
\nn
\eeq
Therefore Condition $(ii)$ follows if the random variables $K^{\alpha,\beta}(X)$ are integrable and we can integrate their moments uniformly on any in finite interval.     
\end{remark}

\begin{remark}
Theorem \ref{thMarginals} can be proved to hold for any approximate SSM $\mS^h=(\pi_0,\M^h,\sg)$ where the kernels $\M^h = \{ \M_1^h, \ldots, \M_K^h \}$ result from an (order 1.0 or better) weak numerical scheme, i.e., {\em not only} for the sequential Euler method.
\end{remark}

%%%%%%%%%%%%%%%%%%%%%%%%%%%%%%%%%%%%%%%
%
%%%%%%%%%%%%%%%%%%%%%%%%%%%%%%%%%%%%%%%
\section{Bayesian filtering} \label{sExamples}

%%%%%%
%
%%%%%%
%\subsection{Optimal and approximate Bayesian filters}

In this section we conduct a numerical study of the performance of a class of discrete-time Bayesian filters which can be used to approximate the posterior laws $\{ \pi_k \}_{k=0}^K$. Such filters rely on a time-discretisation of  $X(t)$ and, therefore, it is of interest to investigate the impact on the filter performance of the sequential Euler scheme introduced in Section \ref{sDiscretisationSDEs}. We outline the filtering algorithms in Section \ref{ssEnKFs} below and then present numerical results in Section \ref{ssL96filters}.

%%%%%%
%
%%%%%%
\subsection{Ensemble Kalman filters} \label{ssEnKFs}

Following \cite{Grudzien20}, we have carried out computer experiments in which we use ensemble Kalman filters (EnKF's) \cite{Evensen03} to assimilate the observations $\{Y_k\}_{k=1}^K$ and approximate the probability laws $\{ \pi_k \}_{k=0}^K$. The EnKF uses a numerical scheme to propagate over time, from $t_{k-1}'$ to $t_k'$, an ensemble of Monte Carlo samples that yield an empirical estimate of the probability distribution of the state. Then, the members of the ensemble undergo a Kalman update for the observation $Y_k$. 

Algorithm \ref{algEnKF} outlines a standard EnKF \cite{Evensen03}. The kernels $\M_k^h$, $k=1, \ldots, K$, used in the prediction stage depend on the numerical scheme employed to approximate the dynamics of the state $X(t)$. For this numerical study, we have implemented EnKF's with the standard Euler scheme the new sequential Euler method of Section \ref{sDiscretisationSDEs}. Each choice of kernel yields a different algorithm, which we label as `Euler EnKF' and `sequential Euler EnKF'.
 
We assume for simplicity that the observations are linear transformations of the states contaminated by Gaussian noise. Specifically, 
\[
Y_k = A_k X(t_k') + Q U_k,
\] 
where $A_k$ is a known $d_y \times d_x$ matrix, $U_k \sim \mN(0,I_{d_y})$ is an i.i.d. sequence of Gaussian vectors and $Q$ is a known matrix parameter. As we cannot generate $X(t_k')$ directly, the filters are implemented under the approximation 
\beq
Y_k \approx A_k \X_k + Q U_k,
\label{eqApproxObs}
\eeq
where $\X_k = X_{n_k}$ is the discrete-time sequence that approximates the state $X(t)$ over the grid $\{t_k'\}_{k=0}^K$, generated by either the standard Euler or the sequential Euler schemes. 

Because of the replacement of the true continuous-time kernels $\M_k^*$ by their discrete-time estimates $\M_k^h$ and the approximate observation equation \eqref{eqApproxObs}, the resulting EnKF algorithm targets the sequence of laws $\{\pi_k^h\}_{k=1}^K$, generated by the approximate SSM $\mS^h$, rather than the true posterior laws $\{\pi_k\}_{k=1}^K$ generated by the exact model $\mS$.

%%%
\begin{algorithm}[htb]
\caption{Ensemble Kalman filter (EnKF) with $M$ samples.}
\label{algEnKF}
\begin{enumerate}
\item \textbf{Initialisation}: generate the initial ensemble by drawing $M$ i.i.d. samples $\X_0^\ii$, $i=1, \ldots, M$, from the prior law $\pi_0$. 
\item \textbf{Recursive step:} for $k = 1, \ldots, K$:
	\begin{enumerate}[a)]
	\item {\em Prediction}: use the discrete-time kernel $\M_k^h$ induced by the numerical scheme to propagate the ensemble from time $t_{k-1}'$ to time $t_k'$. Specifically, generate $M$ new samples
	\beq
	\bar\X_k^\ii \sim \M_k^h(\X_{k-1}^\ii,\sd \X_k), \quad i=1, \ldots, M.
	\nn
	\eeq
	\item {\em Kalman update}: 
		\begin{enumerate}
		\item Propagate the ensemble members through the observation equation,
		\beq
		Y_k^\ii = A_k\bar\X_k^\ii, \quad i=1, \ldots, M.
		\nn
		\eeq 
		Compute the mean vector and the covariance matrix
		\beq
		y_k^M = \frac{1}{M}\sum_{i=1}^M Y_k^\ii, 
		\quad
		C_k^{y,M} = \frac{1}{M-1}\sum_{i=1}^M \left( Y_k^\ii - y_k^M \right)\left( Y_k^\ii - y_k^M \right)^\top,
		\nn
		\eeq
		respectively.
		
		\item Compute the ensemble mean and the cross-covariance matrix
		\beq
		\bar x_k^M = \frac{1}{M}\sum_{i=1}^M \bar\X_k^\ii,
		\quad
		C_k^{xy,M} = \frac{1}{M-1}\sum_{i=1}^M \left(
			\bar\X_k^\ii - \bar x_k^M
		\right)\left(
			Y_k^\ii - y_k^M
		\right)^\top,
		\nn
		\eeq
		respectively.
		
		\item Compute the Kalman gain
		$
		G_k^M = C_k^{xy,M}\left( C_k^{y,M} \right)^{-1}.
		$
		
		\item Update the ensemble,
		\beq
		\X_k^\ii = \bar\X_k^\ii + G_k^M\left( Y_k - Y_k^\ii + Q U_k^\ii \right),
		\quad i=1, \ldots, M,
		\nn
		\eeq
		where $\{ U_k^\ii \}_{i=1}^M$ is a set of $M$ i.i.d. standard Gaussian vectors. 
		\end{enumerate}
	\end{enumerate}
\end{enumerate}
\end{algorithm}
%%%

After the prediction step we can construct an estimate of the one-step-ahead predictive law $\xi_k$ of the form
$
%\beq
\xi_k^{h,M}(\sd x) = \frac{1}{M}\sum_{i=1}^M \delta_{\bar\X_k^\ii}(\sd x),
%\nn
%\eeq
$
while the optimal filter $\pi_k$ is estimated as
$
%\beq
\pi_k^{h,M}(\sd x) =\frac{1}{M}\sum_{i=1}^M \delta_{\X_k^\ii}(\sd x)
%\nn
%\eeq
$
after the update step. Typically, one uses the empirical measures to estimate the posterior mean and covariance matrix of the state. For $\xi_k^{h,M}$, the predictive mean and covariance matrix are $\bar x_k^M$ (in step 2.b.ii of Algorithm \ref{algEnKF}) and $\bar C_k^{h,M} = \frac{1}{M-1}\sum_{i=1}^M \left( \bar\X_k^\ii - \bar x_k^M \right) \left( \bar\X_k^\ii - \bar x_k^M \right)^\top$, respectively. The filtered mean and covariance are computed from $\pi_k^{h,M}$, namely
\beq
x_k^M = \frac{1}{M}\sum_{i=1}^M \X_k^\ii,
		\quad \text{and} \quad
		C_k^{h,M} = \frac{1}{M-1}\sum_{i=1}^M \left(
			\X_k^\ii - x_k^M
		\right)\left(
			\X_k^\ii - x_k^M
		\right)^\top.
		\nn
\eeq

\paragraph{Sequential EnKF} In the standard EnKF of Algorithm \ref{algEnKF} the choice of discrete-time kernel $\M_k^h$ does not affect the Kalman update step directly and, in particular, it does not exploit the structure of the new sequential Euler scheme. Let us assume that the observations $Y_k$ are localised, meaning that we can decompose the $d_y$-dimensional observation vector as
\beq
Y_k = \left[
	\begin{array}{c}
	Y_{0,k}\\
	\vdots\\
	Y_{r-1,k}\\
	\end{array}
\right] = \left[
	\begin{array}{c}
	v_0(X_{j_0}(t_k'))\\
	\vdots\\
	v_{r-1}(X_{j_{r-1}}(t_k'))\\
	\end{array}
\right] + \sigma_yU_k,
\label{eqLocalisedObservations}
\eeq
where the $v_l(\cdot)$'s, $l=0, \ldots, r-1$ are $\Real^{m_x}\mapsto \Real^{m_y}$ observation functions. Each sub-vector $Y_{l,k}$ has dimension $m_y = \frac{d_y}{r}$ and it depends on $X_{j_l}(t_k')$ alone, for some indices $0 \le j_0 \le \ldots \le j_{r-1} < q$. With this localised measurement, it is relatively simple to take advantage of the sequential Euler scheme to design a sequential EnKF (SEnKF) that performs $r$ local Kalman updates (one for each observation $Y_{l,k}$) per time step. 

The SEnKF is outlined in Appendix \ref{apSEnKF}. For simplicity, we assume that $v_l(x)=v_l x$ for some constants $v_l$, $l=0, \ldots, r-1$, and $Q=\sigma_y I_{d_y}$ for some real constant $\sigma_y$, hence $Y_{l,k} = v_l X_{j_l}(t_k') + \sigma_y U_{l,k}$, where $U_{l,k}$ is $\mN(0,I_{m_y})$. The same as in the standard EnKF, we approximate $Y_{l,k} \approx v_l\X_{j_l,k} + \sigma_y U_{l,k}$. Let us remark that the SEnKF algorithm can be applied with nonlinear observation functions $v_l(\cdot)$, as displayed in \eqref{eqLocalisedObservations}, and an arbitrary covariance matrix $\Sigma_y=QQ^\top$ (instead of $\sigma_y^2 I_{d_y}$); however, the notation becomes a bit cumbersome (moreover, note that the covariance of the observation noise can always be diagonalised with a suitable linear transformation).  

The algorithm runs a sequential Euler scheme to propagate an ensemble of $M$ samples $\{ \X_{k-1}^\ii \}_{i=1}^M$ from the time $t_{k-1}' = t_{n_{k-1}}$ of the $(k-1)$-th observation to the time $t_k'=t_{n_k}$ of the $k$-th observation. At time $t_k'$, the procedure generates a predicted ensemble up to the first observation $Y_{0,k}$, namely the state sub-vectors $\{\bar\X_{j_0,k}^\ii\}_{i=1}^M$. This predicted ensemble is then updated via a Kalman gain, to yield $\{\X_{0:j_0,k}^\ii\}_{i=1}^M$ (note that all states $\X_{0,k}^\ii, \ldots, \X_{j_0,k}^\ii$ are updated, not just the $j_0$-th sub-vector). Then the algorithm extends the ensemble up to the coordinate of the second observation, $Y_{1,k}$, and updates the states $\{\X_{0:j_1,k}^\ii\}_{i=1}^M$. This procedure is repeated for each observation up to $Y_{r-1,k}$. Note that the dimension of the updated states increases with each new observation. In particular, the state samples $\{\X_{0:j_0,k}^\ii\}_{i=1}^M$ are updated $r$ times (one per observation), the state samples $\{\X_{j_0+1:j_1,k}^\ii\}_{i=1}^M$ are updated $r-1$ times, etc.

The SEnKF follows rather naturally from the structure of the sequential Euler scheme introduced in Section \ref{ssSTScheme}. However, one can also mimic the SEnKF using the standard Euler scheme for the propagation of the ensemble samples. The resulting filter is very similar and we provide a description in Appendix \ref{apSEnKF2}. Hereafter, we refer to these two algorithms as `sequential Euler SEnKF' and `Euler SEnKF', depending on which numerical scheme they run. 

%%%
\begin{remark}
Neither the `sequential Euler SEnKF' nor the `Euler SEnKF' algorithms are consistent with the (exact) Kalman filter when both the state equation and the observation equation are linear (and the noise terms Gaussian). The reason is that each state variable $\X_{j,k}$, for $j=0, \ldots, d_x-1$, depends on $X_{n_k-1}$ (both in the Euler and the sequential Euler schemes). Hence, for each new observation $Y_{l,k}$, $l=0, \ldots, r-1$, at time $n_k$ one should update the conditional distribution of $X_{n_k-1}$ given the new data. This is not done by these two algorithms, which only update the ensemble members $\X_{0:j_l,k}^{(i)}$.
\end{remark}

%%%%%%
%
%%%%%%
\subsection{Simulation setup: stochastic Lorenz 96 model} \label{ssL96filters}

We have compared numerically the performance of the two versions of the EnKF (`Euler EnKF' and `sequential Euler EnKF') and the two versions of the SEnKF (`Euler SEnKF' and `sequential Euler SEnKF') for the stochastic Lorenz 96 model with linear observations. The signal dynamics are described in Section \ref{sSL96}. The dimension of $X(t)$ is $d_x=200$ and the forcing parameter is $F=8$. 

In our computer experiments observations are collected every $\Delta=0.1$ continuous time units; specifically
\beq 
Y_k = A_k X(t_k') + \sigma_y U_k,
\label{eqObservL96}
\eeq
where $\sigma_y$ is a scale parameter, $U_k \sim \mN(0,I_{d_y})$ are i.i.d. $d_y$-dimensional noise vectors, $t_k' = k\Delta$ for $k=1, \ldots, K$ and $K =\left\lfloor \frac{T}{\Delta} \right\rfloor$. With $T=10$ and $\Delta=0.1$, this yields $K=100$ observation times, with $t_1'=0.1$ and $t_{100}'=10.0$. The observation matrix $A_k$ has dimensions $d_y \times d_x$. It takes the form
\beq
 A_k = \left[
 	e_{m_{0,k}}, e_{m_{1,k}}, \ldots, e_{m_{d_y-1,k}} 
 \right],
 \label{eqAn}
 \eeq
where $e_{m,k}$ is a vector of 0's with a single value of 1 in the $m$-th entry. This observation model implies that
\begin{itemize}
\item at each time $k$ there are $d_y$ state variables which can be observed in Gaussian noise, with $d_y \le d_x$, and
\item the set of observed variables is, in general, different across different observation times.
\end{itemize}
For our simulations, the indices $m_{j,k}$ that determine the state variables to be observed are selected randomly at each time $t_k'$. Specifically, $d_y \le d_x$ indices are drawn from the set $\{0,\ldots,d_x-1\}$ with uniform probabilities and without replacement. Then, they are sorted in ascending order to guarantee that $m_{0,k} < m_{1,k} < \cdots < m_{d_y-1,k}$. The matrix $A_k$ is, therefore, time-varying but it is known to the filtering algorithms at all observation times.

\paragraph{Simulation of a ground-truth signal and observations} Since it is not possible to draw {\em exactly} from the It\^o process $X(t)$, we approximate the ground-truth states $X(t_k')$ for our computer experiments by generating a discrete-time sequence $X_n^o \approx X(t_n^o)$, $t_n^o=nh_o$, by way of the standard Euler scheme with step size $h_o=10^{-5}$. Note that $t_k' = k\Delta = n_kh_o = t_{n_k}^o$ where $n_k = k\frac{\Delta}{h_o}$. This step size $h_o$ is at least two orders of magnitude smaller that the step sizes to be used in the computer experiments for the Markov kernels $\M_k^h$. Given the (approximate) ground truth signal $X_n^o$, we generate synthetic observations of the form 
$Y_k = A_k X_{n_k}^o + \sigma_y U_k$.
The observations simulated in this way are used to construct the estimates $\xi_k^{h,M}$ and $\pi_k^{h,M}$ for different values of $h$ and $M$ depending on the simulation.

%%%%%%
%
%%%%%%
\subsection{Numerical results: robustness}

In the first experiment we assess the capability of the filters to operate with different values of the time step $h$ and the diffusion factor $\sigma$ (see Eq. \eqref{eqL96}). Specifically, we have run simulations with $h \in \{10^{-3}, 5\times 10^{-3}, 10^{-2} \}$ and $\sigma^2 \in \left\{ \frac{1}{4},\frac{1}{2},1,2\right\}$ for $t \in [0,10]$. For each pair $\{h,\sigma^2 \}$, we have run 300 independent simulations and counted how many of them were completed. As shown in Section \ref{sSL96}, a simulation fails to complete when $|X_n|\rw\infty$ (to the machine precision) because of a time step $h$ which is too large. For the filtering algorithms, the sequences $\{ \X_k^\ii \}_{i=1}^M$ depend not only on the dynamics induced by the numerical scheme but also on the updates using the observations $Y_k$. %A filter may fail to complete a simulation when $|\X_k^\ii|\rw\infty$ for {\em some} (not necessarily {\em all}) $i$.     

\begin{figure}[htb]
\centerline{
\subfloat[$h=10^{-3}$; from left to right: $\sigma=\sqrt{1/4}, \sqrt{1/2}, 1$]{
    \label{fPct_a}
    \includegraphics[width=0.32\linewidth]{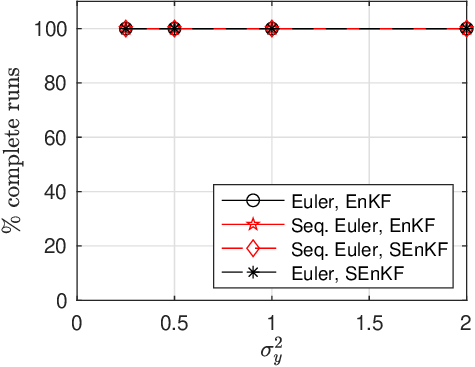}
    \includegraphics[width=0.32\linewidth]{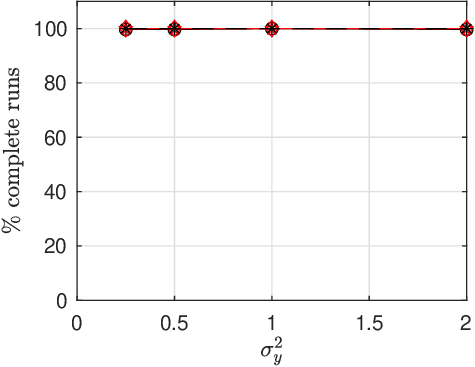}
    \includegraphics[width=0.32\linewidth]{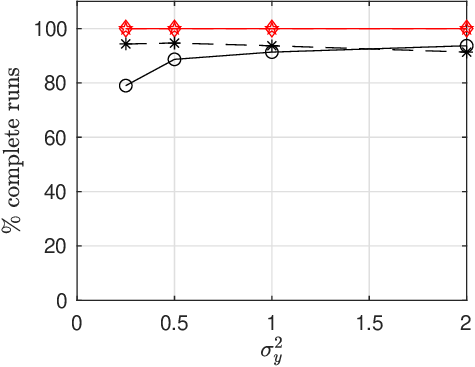}
}
}
\centerline{
\subfloat[$h=5 \times 10^{-3}$; from left to right: $\sigma=\sqrt{1/4}, \sqrt{1/2}, 1$]{
    \label{fPct_b}
    \includegraphics[width=0.32\linewidth]{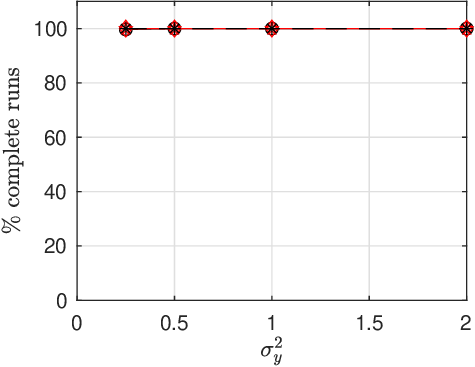}
    \includegraphics[width=0.32\linewidth]{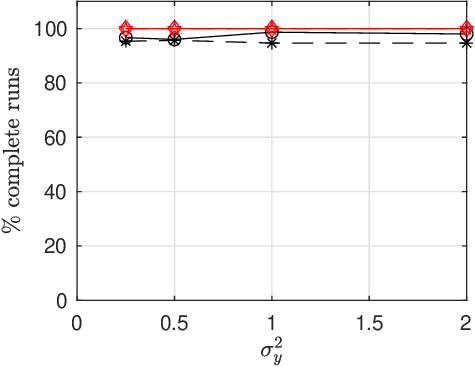}
    \includegraphics[width=0.32\linewidth]{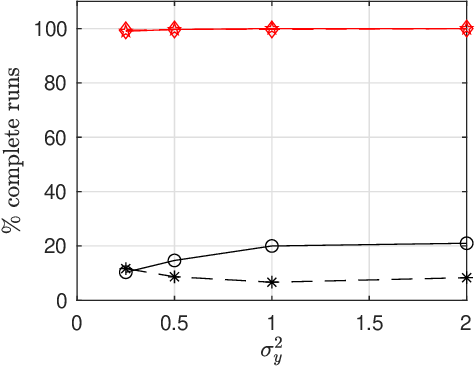}
    
}
}
\centerline{
\subfloat[$h=10^{-2}$; from left to right: $\sigma=\sqrt{1/4}, \sqrt{1/2}, 1$]{
    \label{fPct_c}
    \includegraphics[width=0.32\linewidth]{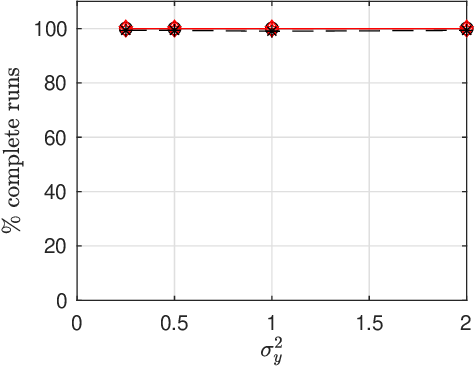}
    \includegraphics[width=0.32\linewidth]{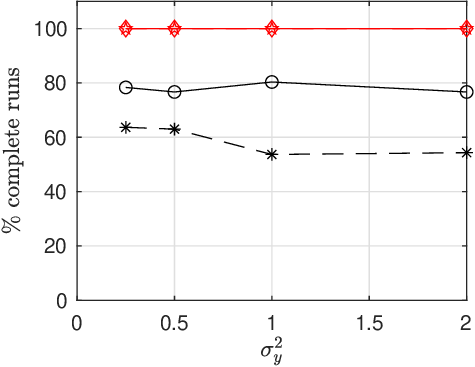}
    \includegraphics[width=0.32\linewidth]{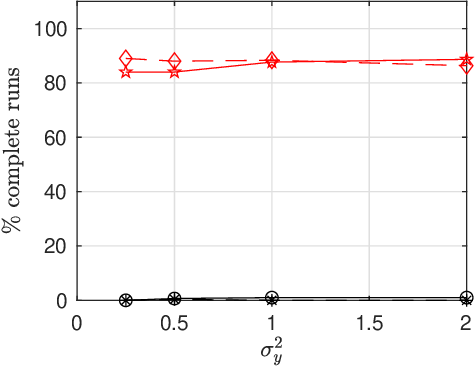}
}
}
\caption{Percentage of complete runs of the EnKF and SEnKF for varying time step $h$ (from top to bottom: $h=10^{-3}, 5 \times 10^{-3}$ and $10^{-2}$) and diffusion factor $\sigma$ (left: $\sigma=\sqrt{1/4}$, middle: $\sigma=\sqrt{1/2}$, right: $\sigma=1$). In each plot, the horizontal axis represents the variance ($\sigma_y^2$) of the observational noise. The length of the simulation is $T=5$ continuous-time units. The continuous-time gap between consecutive observations is $\Delta = 0.1$. The model dimension is $d_x=200$ and the size of the ensemble is $M=d_x=200$. The percentages are estimated from 300 independent simulations.}
\label{fPctEnKF1}
\end{figure}

Figure \ref{fPctEnKF1} displays the estimated percentage of completed simulation runs versus the observation variance ($\sigma_y^2$) for the four versions of the EnKF and each pair $\{h,\sigma\}$. The three plots on the left of Figure \ref{fPctEnKF1} show the percentages of complete simulations for $\sigma=\sqrt{1/4}$ when, from top to bottom, $h= 10^{-3}, 5\times 10^{-3}$ and $10^{-2}$. The plots in the middle and on the right show the percentages of complete runs for the same values of $h$ when $\sigma=\sqrt{1/2}$ and $\sigma=1$, respectively. 

When both $h$ and $\sigma$ are small enough all methods can be used reliably and we have 100\% complete runs for $h=10^{-3}$ and $\sigma \in \{ \sqrt{1/4}, \sqrt{1/2} \}$. However, when the diffusion factor is increased to $\sigma=1$, the filters based on the standard Euler method already suffer a significant percentage of failures, even with $h=10^{-3}$ (see Figure \ref{fPct_a}). When the time step is increased to $h=5 \times 10^{-3}$ (see Figure \ref{fPct_b}), the `Euler EnKF' and `Euler SEnKF' algorithms attain less than 100\% completed simulations already for $\sigma=\sqrt{1/2}$ and suffer a complete breakdown for $\sigma=1$. The `sequential Euler EnKF' and `sequential Euler SEnKF' methods, on the other hand, are fully realiable for $h=5 \times 10^{-3}$, even with $\sigma=1$.

Finally, when $h=10^{-2}$ (see Figure \ref{fPct_c}), the filters based on the new sequential Euler scheme run fully reliably for $\sigma=\sqrt{1/4}$ and $\sigma=\sqrt{1/2}$ and it is only for $\sigma=1$ that they suffer a significant degradation in performance, with a completion rate just over 80\%. The `Euler EnKF' and `Euler SEnKF' algorithms can only be run reliably for $\sigma=\sqrt{1/4}$ and degrade severely already for $\sigma=\sqrt{1/2}$.

\begin{comment}
When the time step is reduced to $h=5\times 10^{-3}$ and $\sigma^2=\frac{1}{2}$, however, the Euler-based filters (Euler EnKF and Euler SEnKF) already fail for $\approx 10\%$ simulation runs, while the filters that rely on the sequential Euler scheme remain robust. The breakdown of Euler-based filters is clear  for $h=10^{-2}$ and $\sigma=\sqrt{1/2}$, while the sequential Euler EnKF/SEnKF algorithms remain reliable.  

Figure \ref{fPctEnKF2} displays the results of the same computer experiments with $\sigma = 1$ (left column of plots) and $\sigma = 2$ (right column of plots). As the intensity of the noise in the state equation is increased, the Euler EnKF and Euler SEnKF algorithms degrade almost completely (with less than 20\% complete runs) in all scenarios except for $h=10^{-3}$ and $\sigma^2=1$, and even in this case the percentage of failures is significant. The sequential Euler EnKF/SEnKF algorithms attain close to 100\% complete runs for $h=10^{-3}$ and the pair $\{ h=5\times 10^{-2}, \sigma^2=1 \}$. In the remaining scenarios the degradation is severe even for these filters.

\end{comment}

Overall, these computer experiments show that the filters (either EnKF or SEnKF) that employ the sequential Euler scheme to approximate the state signal are numerically more robust than their counterparts based on the standard Euler scheme, i.e., they can be reliably applied with smaller values of the time step $h$ (which reduces the computational cost) and in scenarios where the state noise is stronger.

\begin{comment}

\begin{figure}[htb]
\centerline{
\subfloat[$\sigma=1$]{
    \includegraphics[width=0.35\linewidth]{figures_filters/pct_4alg_sx2=1.000_T=5.00_h=0.0010.eps}
}
\subfloat[$\sigma=\sqrt{2}$]{
    \includegraphics[width=0.35\linewidth]{figures_filters/pct_4alg_sx2=2.000_T=5.00_h=0.0010.eps}
}
}
\centerline{
\subfloat[$\sigma=1$]{
    \includegraphics[width=0.35\linewidth]{figures_filters/pct_4alg_sx2=1.000_T=5.00_h=0.0050.eps}
}     
\subfloat[$\sigma=\sqrt{2}$]{
         \includegraphics[width=0.35\linewidth]{figures_filters/pct_4alg_sx2=2.000_T=5.00_h=0.0050.eps}
}
}
\centerline{
\subfloat[$\sigma=1$]{        
         \includegraphics[width=0.35\linewidth]{figures_filters/pct_4alg_sx2=1.000_T=5.00_h=0.0100.eps}
}
\subfloat[$\sigma=\sqrt{2}$]{
         \includegraphics[width=0.35\linewidth]{figures_filters/pct_4alg_sx2=2.000_T=5.00_h=0.0100.eps}
}
}
\caption{Percentage of complete runs of the EnKF and SEnKF for varying time step $h$ (from top to bottom: $h=0.001, 0.005, 0.010$) and diffusion factor $\sigma$ (left: $\sigma=1$, right: $\sigma=\sqrt{2}$). In each plot, the horizontal axis represents the variance ($\sigma_y^2$) of the observational noise. The length of the simulation is $T=5$ continuous-time units. The continuous-time gap between consecutive observations is $\Delta = 0.1$. The model dimension is $d_x=200$ and the size of the ensemble is $M=d_x=200$. The percentages are estimated from 300 independent simulations.}
\label{fPctEnKF2}
\end{figure}

\end{comment}

%%%%%%
%
%%%%%%
\subsection{Numerical results: estimator accuracy}

Next, we assess the performance of the EnKF and the SEnKF for several selected scenarios. By a `scenario', we refer to a combination of values for the diffusion factor $\sigma$ and the observational noise variance $\sigma_y^2$. For each scenario, we select the time step ($h$) to be used by each algorithm in such a way that it is ensured that the simulations can be completed with high probability. For example, for the scenario $\{ \sigma=\sqrt{1/2}, \sigma_y^2 = 1/4 \}$ we can observe in Figure \ref{fPctEnKF1} that the sequential Euler EnKF and the sequential Euler SEnKF algorithms can be run reliably with $h=10^{-2}$, and the Euler EnKF and Euler SEnKF methods are reliable for $h=10^{-3}$. The complete set of time steps for the different filters and scenarios is summarised in Table \ref{tTimeSteps}. Note that the need to operate with different time steps in the numerical scheme in order to guarantee the numerical robustness of the filtering algorithm has an impact on its computational cost (recall Figure \ref{fSchemes}). 

\begin{table}[htb]
\centering
\begin{tabular}{||c|c|c|c|c||}
\hline
$\sigma^2,\sigma_y^2$ 		&Euler 		&Euler 		&Seq. Euler 		&Seq. Euler	\\
							&EnKF  		&SEnKF 		&EnKF				&SEnKF\\
\hline\hline
$\frac{1}{4}, \frac{1}{4}$ 	&$10^{-2}$ 	&$10^{-2}$ 	&$10^{-2}$ 			&$10^{-2}$ 	\\
\hline
$\frac{1}{2}, \frac{1}{4}$ 	&$10^{-3}$ 	&$10^{-3}$ 	&$10^{-2}$ 			&$10^{-2}$ 	\\
\hline
$1, 1$ 						&$10^{-4}$ 	&$10^{-4}$ 	&$5\times 10^{-3}$ 	&$5\times 10^{-3}$ \\
\hline
%$2,1$ 						&$10^{-4}$ 	&$10^{-4}$ 	&$10^{-3}$ 			&$10^{-3}$ \\
%\hline
\end{tabular}
\caption{Time steps ($h$) used for the different discrete-time filters in Figure \ref{fScenarios}.}
\label{tTimeSteps}
\end{table}   

The performance of the filters is compared in terms of their normalised mean square error (NMSE). To be explicit, let $X_n^o$ be the approximate `ground-truth' signal generated using the Euler scheme with $h_o=10^{-5}$ and let $\X_k^o = X_m^o$, where $m=\frac{k\Delta}{h_o}$, be the ground truth signal at the observation time $t_k'$. We also denote the estimate of $\X_k^o$ computed via an EnKF or SEnKF algorithm with ensemble of size $M$ as $\X_k^M$. Then, we define the NMSE as
\beq
\text{NMSE}_M := \frac{
	\sum_{k=1}^K \| \X_k^o - \X_k^M \|^2
}{
	\sum_{k=1}^K \| \X_k^o \|^2
}.
\nn
\eeq
If $\X_k^o$ represents a physical magnitude, then the NMSE is the power of the error normalised by the power of the signal of interest. 

For each scenario in Table \ref{tTimeSteps} we have carried out 40 independent simulations. In each one of them, we run all five filtering algorithms for the same ground truth and observations, with increasing value of the ensemble size, namely $M=50, 100, 200, 300$ and $400$, and compute the resulting values of NMSE$_M$. These errors are then averaged over the set of 40 simulation trials.

\begin{figure}[htb]
\centerline{
    \subfloat[$\sigma^2=\frac{1}{4}$, $\sigma_y^2=\frac{1}{4}$]{
        \label{fScenarios_a}
        \includegraphics[width=0.35\textwidth]{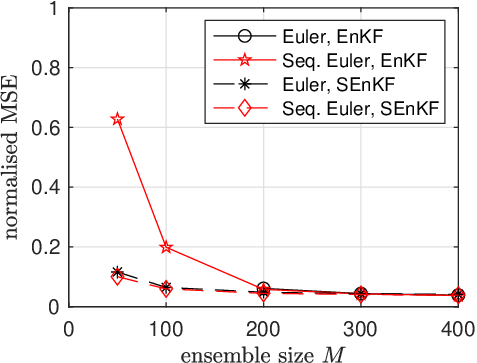}
        \includegraphics[width=0.35\textwidth]{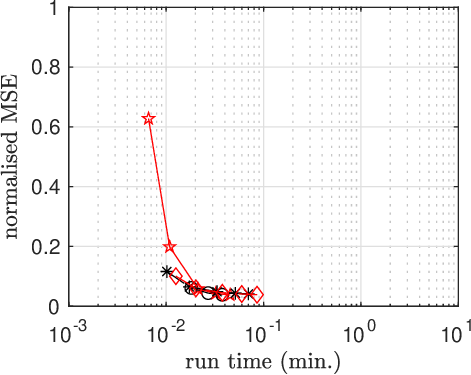}
    }
}
\centerline{
    \subfloat[$\sigma^2=\frac{1}{2}$, $\sigma_y^2=\frac{1}{4}$]{
        \label{fScenarios_b}
        \includegraphics[width=0.35\textwidth]{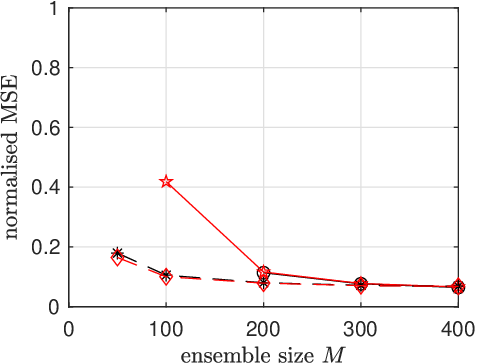}
        \includegraphics[width=0.35\textwidth]{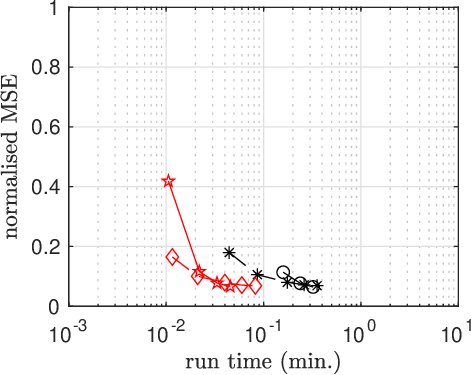}
    }
}
\centerline{
    \subfloat[$\sigma^2=1$, $\sigma_y^2=1$]{
        \label{fScenarios_c}
        \includegraphics[width=0.35\textwidth]{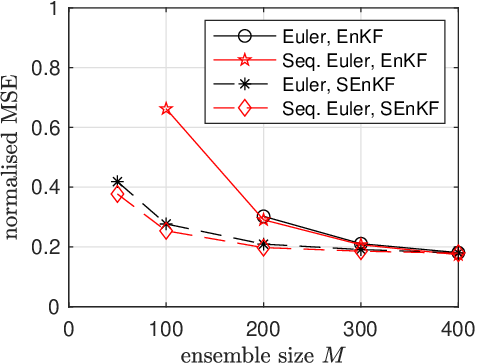}
        \includegraphics[width=0.35\textwidth]{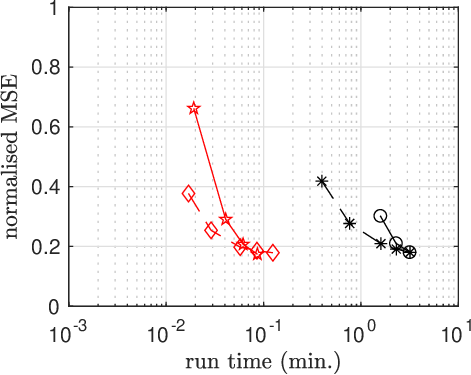}
    }
}
\caption{\textbf{Left:} NMSE vs. ensemble size $M$ for the Euler EnKF, Euler SEnKF, sequential Euler EnKF and sequential Euler SEnKF algorithms in three different scenarios. \textbf{Right:} NMSE vs. run-time for the same set of simulations.}
\label{fScenarios}
\end{figure}

Figure \ref{fScenarios} displays the results. The plots on the left show the averaged NMSE's versus the ensemble size $M$, while the plots on the right show, for the same set of computer experiments, the NMSE's versus the run-time of each algorithm. Note that Figure \ref{fScenarios} does not display an error value for each value of $M$ in every scenario. For example, in Figure \ref{fScenarios_b} $\{ \sigma=\sqrt{1/2}, \sigma_y^2=1/4 \}$, there are no NMSEs for the Euler EnKF method with $M<200$. This means that, for this algorithm, some simulations did not complete despite the careful choice of $h$ (the results in Figure \ref{fPctEnKF1}, from which the time steps are selected, have been obtained with $M=d_x=200$).

If we observe the plots on the left of Figure \ref{fScenarios}, we see that 
the estimation error decreases as $M$ increases and, for $M \ge 200$, all algorithms attain a similar accuracy. For $M<200$, however, the SEnKF-based methods are more accurate and numerically more robust than the EnKF-based algorithms. %When the diffusion factor is increased ($\sigma=\sqrt{2}$ in Figure \ref{fScenarios2_b}) the filters built around the standard Euler scheme, including the SEnKF algorithm, fail for all values of $M$. The sequential Euler SEnKF is the only algorithm that runs for all values of $M$ in Figure \ref{fScenarios2_b} and it also attains a smaller error then the sequential Euler EnKF for $M \ge 200$ in this case.
The Euler EnKF, in particular, does not run reliably with $M< 200$.

The plots on the right of Figure \ref{fScenarios} display the NMSE's of the filters versus their run-time. Note that the differences in computational cost for different filters and same $M$ are due to:
\begin{itemize}
\item the choice of numerical scheme (the sequential Euler scheme is heavier than the standard Euler method),
\item the choice of $h$ (a smaller time step implies a larger number of discrete time steps to be taken by the filters), and
\item the choice of EnKF or SEnKF technique (the sequential processing of the observations at a given time $t_k'$ demands additional computations).
\end{itemize}
For Figure \ref{fScenarios_a} (`small' dynamical and observational noise) all algorithms attain a similar performance, except that the sequential Euler EnKF performs clearly worse for $M=50$. Note that in this scenario all filters operate with $h=10^{-2}$.

In Figs. \ref{fScenarios_b} and \ref{fScenarios_c} we see that the algorithms based on the standard Euler scheme become less efficient because they require very small time steps to run reliably. %For \ref{fScenarios2_b}, both the Euler EnKF and the Euler SEnKF methods break down even with $h=10^{-4}$. 
Also, the filters based on the sequential Euler scheme become relatively more efficient, compared to Euler EnKF and Euler SEnKF, as the diffusion factor $\sigma$ is increased. In the scenario of Figure \ref{fScenarios_c}, the filters based on the sequential Euler scheme perform clearly better (they demand a smaller run-time for the same accuracy).

%%%%%%%%%%%%%%%%%%%%%%%%%%%%%%%%%%%%%%%
%
%%%%%%%%%%%%%%%%%%%%%%%%%%%%%%%%%%%%%%%
\section{Summary and outlook} \label{sConclusions}

%%%
%
%%%
\subsection{Summary}

We have introduced a new predictor-corrector numerical scheme that can be applied to the time-discretisation of a broad class of multivariate It\^o SDEs. The key feature of the new method is that it operates sequentially and recursively along the dimensions of the It\^o process. We have shown through computer experiments, using the stochastic Lorenz 96 system as a test model, that this feature makes the new scheme numerically robust, in the sense that it can operate on coarser time grids than other (similar) schemes. This is advantageous for the approximation of large-dimensional processes because it reduces the number of discrete-time samples that have to be generated and stored.  

The specific algorithm that we have investigated is derived from the backward Euler method and we have proved that it attains weak order 1.0. However, the same strategy can be applied to obtain sequential and recursive schemes starting from other implicit methods, possibly of higher order.   

The second half of the paper has been devoted to the application of the new scheme in the context of Bayesian filtering. We have considered a class of state space models where the state dynamics are modelled by an It\^o SDE and the observations are collected instantaneously, over a given time grid. The goal of Bayesian filtering algorithms is to compute the probability law of the state at each observation time, conditional on the available observations up to that time. Exact solutions are not tractable in general and a common approach is to approximate the dynamics of the state using a time discretisation scheme. In this way, one obtains a simpler discrete-time state space model. A major contribution of the paper is to prove that, under some regularity assumptions, the marginal probability laws generated by the approximate discrete-time model converge to the laws generated by the original continuous-time model. This result holds for the new sequential Euler discretisation scheme, but also for any other weakly-convergent scheme.     

Finally, we have assessed the performance of several discrete-time ensemble Kalman filters that incorporate the proposed sequential Euler scheme and the standard Euler-Maruyama method. The computer experiments have shown that the filters employing the new sequential scheme can operate with larger time steps, smaller Monte Carlo ensembles and noisier systems. As a result, they are more efficient, attaining the same accuracy as their standard Euler counter-parts with a smaller computational cost. 

%%%
%
%%%
\subsection{Outlook}

We envisage the continuation of this research in several directions:
\paragraph{SEnKF algorithm} The SEnKF algorithm incorporates the sequential predictor-corrector scheme in a very natural fashion and we have attained good performance in our computer experiments with the stochastic Lorenz 96 model. However, further research is needed for this algorithm in order to assess, e.g., its performance with nonlinear observations and/or correlated observational noise, as well as to quantify the bias with respect to an exact Kalman filter when the SSM of interest is linear.
\paragraph{Particle filtering} In this paper, we have chosen the family of EnKF's to illustrate the design of Bayesian filters based on the sequential predictor-corrector Euler scheme. This sequential discretisation scheme can also be naturally combined with the family of particle filtering methods. These algorithms can exploit the sequential structure of the discretisation method and, compared to the SEnKF, they enjoy theoretical guarantees of consistency with the optimal discrete-time filter.
\paragraph{Sequential predictor-corrector Runge-Kutta scheme} Other methods beyond the backward Euler can be made sequential using the proposed predictor-corrector approach. In particular, we conjecture that significant performance improvements may be attained if the new methodology is applied to Runge-Kutta (RK) schemes. In particular, the explicit RK scheme of \cite{Ruemelin82} has been shown to attain good performance (in terms of accuracy versus run-time) in \cite{Grudzien20}. We believe that the investigation of a sequential version of this RK scheme is of  interest.
\paragraph{Application to stochastic partial differential equations (SPDEs)} SPDEs are often reduced to systems of SDEs as a prior step to their time discretisation \cite{Lord14}. As a consequence, it is also possible to apply the proposed approach to these models. In particular, finite-difference schemes for SPDEs present significant opportunities for improvement since the spatial domain can be sequentially covered in many ways --with some of them potentially better than others.
\paragraph{Multi-level Monte Carlo} Another family of methods that can benefit from the interplay with sequential predictor-corrector schemes is the class of multi-level Monte Carlo (MLMC) algorithms \cite{hoel2016multilevel,beskos2017multilevel,jasra2017multilevel,jasra2020multilevel}. These methods build multiple estimators at different accuracy levels and combine them to obtain a final estimator with a telescoping sum. In high-dimensional models, sequential predictor-corrector discretisation schemes can be embedded into multilevel filtering methods to improve estimation at each level, see, e.g., \cite{hoel2016multilevel} for multilevel EnKF methods. Moreover, sequential discretisation can also be combined with particle filtering, as mentioned above, hence multilevel extensions of these methods \cite{jasra2017multilevel,jasra2020multilevel} can be explored as well.

%%%%%%%%%%%%%%%%%%%%%%%%%%%%%%%%%%%%%%%
%															%
%%%%%%%%%%%%%%%%%%%%%%%%%%%%%%%%%%%%%%%
\appendix

%%%%%%%%%
%
%%%%%%%%%
\section{Proof of Theorem \ref{thWeak}} \label{apProofThWeak}
Before we proceed with the proof of Theorem \ref{thWeak}, let us introduce the continuous-time approximation 
\begin{equation}
\bar X\left( t\right) = \left[
	\begin{array}{c}
	\bar X_0(t)\\
	\vdots\\ 
	\bar X_{q-1}(t)\\
	\end{array} 
\right]
\label{eqBarX}
\end{equation}
constructed as 
\begin{equation}
\bar X_i(t) = X_i(0) + \int_0^t f_i\left( \tilde X^{i,h}, p \right) \sd p + \int_0^t s_i\left( \bar X, \tau_p \right) \sd W_i(p).  
\label{eqBarXi}
\end{equation}
where $\tau_p = \left\lfloor \frac{p}{h} \right\rfloor h$ and the $d_x$-dimensional vectors 
$
\tilde X^{i,h}(p) = \left[ 
	\begin{array}{c}
	\tilde X_0^{i,h}(p)\\
	\vdots\\
	\tilde X_{q-1}^{i,h}(p)\\
	\end{array}
\right]
$ are defined as 
\beq
\tilde X^{i,h}_j(p) =\left\{ 
	\begin{array}{cl}
	\bar X_j\left( \left\lceil \frac{p}{h} \right\rceil h \right),  &\text{if $0 \le j < i$}, \\ 
	&\\
	\bar X_j\left(  \left\lfloor \frac{p}{h} \right\rfloor h \right) + hf_i\left(\bar 	X_j\left( \left\lfloor \frac{p}{h} \right\rfloor h \right), \left\lfloor \frac{p}{h} \right\rfloor h \right)
	&\text{if $i \le j < q$}\\
	\end{array}%
\right.,
\nn
\eeq
for $i=0,...,q-1$. Note that the signals $\bar X(t)$ and $\tilde X^{i,h}(t)$ are estimates of $X(t)$ and, hence, they are $d_x$-dimensional. Their components $\bar X_j(t)$ and $\tilde X^{i,h}_j(t)$, with $j=0, \ldots, q-1$, are $m_x \times 1$ vectors. An induction argument shows that $\bar X\left( nh\right) = X_n$, for $n=0, 1, \ldots, N$.

Additionally, let us introduce the family of $\sigma$-algebras $\{\mF_{n,i}: n=0, \ldots, N; ~~ i=0, \ldots, d_x-1 \}$ such that
	\begin{itemize}
	\item $\mF_{n,i} \subseteq \mF_{m,j}$ whenever $n < m$,
	\item $\mF_{n,i} \subseteq \mF_{n,j}$ whenever $i \le j$, and
	\item the first $i$ entries of $\bar X(nh)$, denoted $\bar X^{0:i-1}(nh)$, are measurable w.r.t. $\mF_{n,i-1}$.
	%\item $\tilde X^{i,h}(t)$ is measurable w.r.t. $\mF_{n,j}$ for $0 \le t < (n+1)h$ and $i \le j$.
	\end{itemize}
The $\sigma$-algebra $\mF_{n,i}$ can be generated by the initial condition $\bar X(0)$, the $d_x$-dimensional Wiener process $W(t)$ for $0 \le t \le (n-1)h$ and $W_{0:i-1}(t)$ for $(n-1)h < t \le nh$.

%%%
\begin{proof} 
The argument below is a variation of the procedure in the proof of Theorem 14.1.5 in \cite{Kloeden95}. Let $L^0$ denote the operator 
\beq
L^{0} = \partial_t 
	+ \sum_{i=0}^{d_{x}-1} f^i \partial_{x_i} 
	+ \frac{1}{2} \sum_{i=0}^{d_x-1} \sum_{j=0}^{d_x-1} s^{i,\cdot} s^{\cdot,j}\partial_{x_i,x_j},
\nn
\eeq
where $s^{i,\cdot}$ and $s^{\cdot,j}$ are the $i$-th row and $j$-th column, respectively, of the diffusion coefficient $s$, and for some map $v:\Real^{d_x} \times [0,T] \mapsto \Real$, we denote $\partial_t v(x,t) = \frac{\partial v(x,t)}{\partial t}$, $\partial_{x_i}v(x,t) = \frac{ \partial v(x,t)}{\partial x_i}$ and $\partial_{x_i,x_j} v(x,t) = \frac{\partial^2 v(x,t)}{\partial x_i \partial x_j}$. From  \cite[Theorem 4.8.6]{Kloeden95} we know that the functional 
\beq
v(x,t) := \mbE\left[ \phi(X(T)) | X(t)=x \right], \quad
t \in [0,T], \quad x \in \Real^{d_x}, \label{eqDefV}
\eeq 
is a solution of the final value problem
\beq
L^0 v = 0, \quad \text{with} \quad v(x,T) = \phi(x). \label{eqPDE}
\eeq
Moreover, $v(x,t)$ is four times continuously differentiable in the argument $x = [x_0, \ldots, x_{d_x-1}]^\top$ and these partial derivatives are uniformly bounded, hence $v(\cdot,t) \in C_B^4(\Real^{d_x})$. 

From \eqref{eqPDE} and It\^o's formula we readily obtain that $\mbE\left[ v(X(t),t) \right] = \mbE\left[ v(X(0),0) \right]$ for any $t \in [0,T]$ and, since $X(0) = X_0$, we have the identity
\beq
\phi\left( X(T) \right) = \mbE\left[ v(X(T),T) \right] = \mbE\left[ v(X_0,0) \right].
\eeq
From Eq. \eqref{eqDefV}, it also follows that $\phi(X_N)=\mbE\left[ v(X_N,T) \right]$. Since $X_N=\bar X(T)$ and $\bar X(0) = X_0$, we finally obtain the relationship
\beqa
\left|
	\mbE\left[
		\phi\left( 
			X(T) 
		\right)
	\right] - \mbE\left[
		\phi(X_N)
	\right]
\right| &=& \left|
	\mbE\left[ v(X(0),0) \right] - \mbE\left[ v(\bar X(T), T) \right]
\right| \nn\\
&=& \left|
	\mbE\left[ v(\bar X(0),0) \right] - \mbE\left[ v(\bar X(T),T) \right]
\right| \nn\\
&=& \left|
	\mbE\left[
		v(\bar X(T),T) - v(X_0,0)
	\right]
\right|. \nn
\eeqa
Hence, we have rewritten the error $\left|
	\mbE\left[
		\phi\left( 
			X(T) 
		\right)
	\right] - \mbE\left[
		\phi(X_N)
	\right]
\right|$
in terms of the signal $\bar X(t)$ alone, which has been constructed to satisfy $\bar X(nh) = X_n$ for $n=0, \ldots, N$, and $Nh=T$.

Using It\^o's formula for the process $\bar X(t)$, we readily obtain 
\beqa
v(\bar X(T),T) - v(X_0, 0) &=& \int_0^T \left[
	\partial_t v\left( \bar X(u),u \right) 	
	+ \sum_{i=0}^{d_x-1} f^i\left(
		\tilde X^{\lfloor i/m_x \rfloor,h}, u
	\right)\partial_{x_i} v\left( \bar X(u), u \right) 
\right. \nn\\
&& \quad \quad \left.
	+ \frac{1}{2}\sum_{i=0}^{d_x-1}\sum_{j=0}^{d_x-1} s^{i,\cdot}(\bar X, \tau_u) s^{\cdot,j} (\bar X, \tau_u) \partial_{x_i,x_j}v\left( \bar X(u), u \right)
\right] \sd u \nn\\
&& + \int_0^T \sum_{i=0}^{d_x-1}  s^{i,\cdot}(\bar X, \tau_u) \partial_{x_i} v\left( \bar X(u), u \right) \sd W(u)  
\label{eqItoXbar}
\eeqa
while, using \eqref{eqPDE},
\beqa
0 = \int_0^T L^0 v(\bar X(u),u)\sd u &=& \int_0^T \left[
	\partial_t v\left( \bar X(u),u \right) 	
	+ \sum_{i=0}^{d_x-1} f^i\left(
		\bar X, u
	\right)\partial_{x_i} v\left( \bar X(u), u \right) 
\right. \nn\\
&& \quad \quad \left.
	+ \frac{1}{2}\sum_{i=0}^{d_x-1}\sum_{j=0}^{d_x-1} s^{i,\cdot}(\bar X, u) s^{\cdot,j} (\bar X, u) \partial_{x_i,x_j}v\left( \bar X(u), u \right)
\right] \sd u. \label{eqL0v=0}
\eeqa
Combining Eqs. \eqref{eqItoXbar} and \eqref{eqL0v=0} and taking expectations yields
\beqa
&&\mbE\left[
	v(\bar X(T),T) - v(X_0, 0)
\right] =  \nn\\
&&\mbE\left\{
	\int_0^T \sum_{i=0}^{d_x-1} \left[
		f^i\left(
			\tilde X^{\lfloor i/m_x \rfloor,h}, u
		\right) - f^i\left(
			\bar X, u
		\right) 
	\right] \partial_{x_i} v\left( \bar X(u), u \right) \sd u
\right\}  \nn \\
&&+ \frac{1}{2} \mbE\left\{
	\int_0^T \sum_{i=0}^{d_x-1}\sum_{j=0}^{d_x-1} \left[
		s^{i,\cdot}(\bar X, \tau_u) s^{\cdot,j} (\bar X, \tau_u)
		- s^{i,\cdot}(\bar X, u) s^{\cdot,j} (\bar X, u)
	\right] \partial_{x_i,x_j}v\left( \bar X(u), u \right) \sd u
\right\}, \nn\\
\label{eqBigEq}
\eeqa
as 
$
\mbE\left[
	\int_0^T \sum_{i=0}^{d_x-1} s^{i,\cdot}(\bar X, \tau_u) \partial_{x_i} v\left( \bar X(u), u \right) \sd W(u)
\right] = 0
$.

Let $\sff$ denote either $\sff = f^i \partial_{x_i} v$ or $\sff = \partial_{x_i} v$, for $i=0, \ldots, d_x-1$. Using the uniform bound $B<\infty$ on the derivatives of $f^i$ and $v$ and the fact that $\| \bar X(u) - \tilde X^{\lfloor i/m_x \rfloor,h}(u) \| = \mO(h)$ it is straightforward to prove that
\beq
\left| 
	\mbE\left[ 
		\sff\left( \tilde X^{\lfloor i/m_x \rfloor,h}(u), u \right) 
		- \sff\left( \bar X(u), u \right)
	\left\vert \mF_{\tau_u,i} \right. \right] 
\right| \leq c_{1,\sff} h,
\label{eqCst1}
\eeq
where $c_{1,\sff}<\infty$ is independent of $h$, the initial condition $X_0$ and the coordinate $i$. Similarly, when either $\sff = s^{i,\cdot}s^{\cdot,j}\partial_{x_i,x_j} v$ or $\sff=\partial_{x_i,x_j} v$ one can prove that
\beq
\left| 
	\mbE\left[ 
		\sff\left( \bar X(\tau_u), \tau_u \right) 
		- \sff\left( \bar X(u), u \right)
	\left\vert \mF_{\tau_u-h,d_x-1} \right. \right] 
\right| \leq c_{2,\sff} h,
\label{eqCst2}
\eeq
where $c_{2,\sff}<\infty$ is independent of $h$, $i$ and $X_0$. We denote $c_\sff = c_{1,\sff} \vee c_{2,\sff} < \infty$.

Combining \eqref{eqBigEq} with the inequalities \eqref{eqCst1} and \eqref{eqCst2} we can obtain a suitable upper bound for 
$\left| 
	\mbE\left[
		v(\bar X(T),T) - v(X_0, 0)
	\right]
\right|
$. To be specific, for the first term on the right-hand side of \eqref{eqBigEq} we obtain
\beqa
\left|
	\mbE\left\{
		\int_0^T \sum_{i=0}^{d_x-1} \left[
			f^i\left(
				\tilde X^{\lfloor i/m_x \rfloor,h}, u
			\right) - f^i\left(
				\bar X, u
			\right) 
		\right] \partial_{x_i} v\left( \bar X(u), u \right) \sd u
	\right\}
\right| &\le& \nn \\
\mbE\left\{
	\int_0^T \sum_{i=0}^{d_x-1} \left|
		\mbE\left[
			f^i\left(
				\tilde X^{\lfloor i/m_x \rfloor,h}, u
			\right) \partial_{x_i}v\left( \tilde X^{\lfloor i/m_x \rfloor,h}, u \right)
			\right.
        \right.
    \right.
&& \nn\\
\left.
    \left.
        \left.
            - f^i\left(
				\bar X, u
			\right) \partial_{x_i} v\left( \bar X(u), u \right)
			\left\vert \mF_{\tau_u,i} \right.
		\vphantom{\sum} \right]
	\right| \sd u
\vphantom{\int}\right\} && \nn \\
+ \mbE\left\{
	\int_0^T \sum_{i=0}^{d_x-1} \left|
		f^i\left(
				\tilde X^{\lfloor i/m_x \rfloor,h}, u
			\right)
	\right| \left|
		\mbE\left\{
			\left[\delta
				\partial_{x_i} v
			\right]
			\left\vert \mF_{\tau_u,i} \right.
		\right\}
	\right| \sd u
\right\} &\le&\nn \\
T d_x c_\sff h + \mbE\left\{
	\int_0^T \sum_{i=0}^{d_x-1}	\mbE\left[
	    \left|
			f^i\left(
				\tilde X^{\lfloor i/m_x \rfloor,h}, u
			\right)
		\right| \left|
			\delta \partial_{x_i} v \right|\right] \sd u
\right\} &\le&\nn \\
T d_x c_\sff h + c_\sff h \mbE\left[
	\int_0^T \sum_{i=0}^{d_x-1} \left|
		f^i\left(
			\tilde X^{\lfloor i/m_x \rfloor,h}, u
		\right)
	\right|
\right] &\le& \nn \\
T d_x c_\sff \left( 
	1 + \sup_{0 \le u \le T} \mbE\left[
		\left|
			f^i\left(
				\tilde X^{\lfloor i/m_x \rfloor,h}, u
			\right)
		\right|
	\right]
\right) h &\le& C' h,\nn    
\eeqa
where we have denoted
$
\delta\partial_{x_i} v := \partial_{x_i} v\left( \bar X(u), u \right) - \partial_{x_i}v\left( \tilde X^{\lfloor i/m_x \rfloor,h}, u \right)
$. In the sequence of bounds above, the second inequality follows from \eqref{eqCst1}, the third inequality follows from \eqref{eqCst2} and we obtain the fourth inequality from the uniform bound on $f^i$; hence, $C' = \mO(T d_x)$ is independent of $h$ and $X_0$. 

By a similar argument, for the second term on the right-hand side of \eqref{eqBigEq} there exists a constant $C'' = \mO(Td_x^2) <\infty$, independent of $h$ and $X_0$, 
such that
\beq
\frac{1}{2} \mbE\left\{
	\int_0^T \sum_{i=0}^{d_x-1}\sum_{j=0}^{d_x-1}  \delta s^{i,\cdot}\partial_{x_i,x_j}v\left( \bar X(u), u \right) \sd u
\right\} \le C'' h.
\eeq
where
$
\delta s^{i,\cdot} :=
		s^{i,\cdot}(\bar X, \tau_u) s^{\cdot,j} (\bar X, \tau_u)
		- s^{i,\cdot}(\bar X, u) s^{\cdot,j} (\bar X, u)
$.
If we set $C=C'+C''<\infty$, then
\beq 
\left|
	\phi(X(T))-\phi(X_N)
\right| = \mbE\left[
	v(\bar X(T),T) - v(\bar X(0), 0)
\right] \le C h,
\eeq
where $C=\mO(Td_x^2)$ is independent of $h$ and $X_0$. 
\end{proof}

%%%%%%%%%%
%
%%%%%%%%%%
\section{Proof of Theorem \ref{thMarginals}} \label{apProofThMarginals}
We follow an induction argument. By construction, the exact and approximate SSMs share the same prior law $\pi_0$, hence $\pi_0=\pi_0^h$. Let us now assume that
\beq
D(\pi_{k-1},\pi^h_{k-1}) \le C_{k-1} h
\label{eqP10}
\eeq
for some $1 \le k < K$ and let $\phi \in C_B^4(\Real^{d_x})$ be a test function such that $\| \phi \|_\infty \le 1$. We readily see that
\beq
| \xi^h_k(\phi) - \xi_k(\phi) | = \left| \M_k^h\pi^h_{k-1}(\phi) - \M_k^*\pi_{k-1}(\phi) \right| 
= \left|
	\pi^h_{k-1}( \bar\phi_k^h ) - \pi_{k-1}( \bar\phi_k )
\right|, \label{eqP11}
\eeq
where 
\beq
\bar \phi_k(x_{k-1}) = \int \phi(x_k) \M_k^*(x_{k-1},\sd x_k)
\quad \text{and} \quad
\bar \phi^h_k(x_{k-1}) = \int \phi(x_k) \M_k^h(x_{k-1},\sd x_k), \nn
\eeq
hence a triangle inequality yields
\beq
\left|
	\pi^h_{k-1}( \bar\phi_k^h ) - \pi_{k-1}( \bar\phi_k )
\right| \le \left| 
	\pi^h_{k-1}(\bar\phi_k^h) - \bar \pi^h_{k-1}(\bar\phi_k) 
\right| + \left|
	\bar \pi^h_{k-1}(\bar\phi_k) - \pi_{k-1}( \bar\phi_k )
\right|.
\label{eqP12}
\eeq

It is straightforward to show that $\|\phi\|_\infty \le 1$ implies $\| \bar\phi_k \|_\infty \le 1$. Moreover, assumption (ii) in the statement of Theorem \ref{thMarginals} implies that 
\beq
\sup_{k\le K, |\alpha|\le 4} \| \bar \phi_k^{(\alpha)} \|_\infty < B';
\nn
\eeq 
hence, $\bar \phi_k \in C_B^4(\Real^{d_x})$. Therefore, we can apply the induction hypothesis \eqref{eqP10} to obtain
\beq
\sup_{\| \bar{\phi}\|_\infty \le 1} \left|
	\bar \pi^h_{k-1}(\bar\phi_k) - \pi_{k-1}( \bar\phi_k )
\right| = C_{k-1}h,
\label{eqP12.5}
\eeq
which accounts for the second term on the right-hand side of \eqref{eqP12}. For the first term on the right-hand side of \eqref{eqP12}, we note that Lemma \ref{lmConvK} yields
\beq
\sup_{x\in \Real^{d_x}} | \bar\phi^h_k(x) - \bar\phi_k(x) | \le Ch
\nn
\eeq
%and, since $\sup_k \| \bar\phi^h_k \|_\infty \le 1$, the dominated convergence theorem yields
and, therefore,
\beq
\left| 
	\pi^h_{k-1}(\bar\phi_k^h) - \bar \pi^h_{k-1}(\bar\phi_k) 
\right| = \left| 
	\pi^h_{k-1}( \bar\phi_k^h - \bar\phi_k ) 
\right| \le Ch.
\label{eqP13}
\eeq
Substituting \eqref{eqP13} and \eqref{eqP12.5} in \eqref{eqP12}, and then \eqref{eqP12} back in \eqref{eqP11}, yields
\beq
D(\xi_k^h,\xi_k) \le \bar C_k h,
\label{eqP14}
\eeq
where $\bar C_k = C_{k-1} + C<\infty$.

Next, we write the difference $\pi^h_k(\phi) - \pi_k(\phi)$ in terms of $\xi^h_k$ and $\xi_k$ as
\beqa
\pi^h_k(\phi) - \pi_k(\phi) &=& \frac{
	\xi_k^h(g_k\phi)
}{
	\xi_k^h(g_k)
} - \frac{
	\xi_k(g_k\phi)
}{
	\xi_k(g_k)
} \pm \frac{
	\xi_k^h(g_k\phi)
}{
	\xi_k(g_k)
} \nn\\
&=& \frac{
	\xi_k^h(g_k\phi) - \xi_k(g_k\phi)
}{
	\xi_k(g_k)
} + \frac{
	\xi_k^h(g_k\phi)
}{
	\xi_k^h(g_k)
} \frac{
	\xi_k(g_k) - \xi_k^h(g_k)
}{
	\xi_k(g_k)
}, \nn
\eeqa
which readily yields the bound
\beq
\left| \pi^h_k(\phi) - \pi_k(\phi) \right| \le
\frac{
	\left|
		\xi_k^h(g_k\phi) - \xi_k(g_k\phi)
	\right|
}{
	\xi_k(g_k)
} + \| \phi \|_\infty \frac{
	\left|
		\xi_k(g_k) - \xi_k^h(g_k)
	\right|
}{
	\xi_k(g_k)
}.
\label{eqP15}
\eeq
The difference $\left|
	\xi_k^h(g_k\phi) - \xi_k(g_k\phi)
\right|$ can be upper bounded as 
\beq 
\left|
	\xi_k^h(g_k\phi) - \xi_k(g_k\phi)
\right| \le \|\phi\|_\infty \left|
	\xi_k(g_k) - \xi_k^h(g_k)
\right|
\label{eqP15.5}
\eeq
and, since $g_k \in C_B^4(\Real^{d_x})$ and $\|g_k\|_\infty \le 1$, the inequality \eqref{eqP14} implies that
\beq
\left|
	\xi_k(g_k) - \xi_k^h(g_k)
\right| \le \bar C_k h.
\label{eqP16}
\eeq
Substituting \eqref{eqP16} and \eqref{eqP15.5} into \eqref{eqP15} we obtain the bound
\beq
\left| \pi^h_k(\phi) - \pi_k(\phi) \right| \le \frac{
	2 \|\phi\|_\infty \bar C_k h
}{
	\xi_k(g_k)
} \label{eqP17}
\eeq
which holds for all $\phi \in C_B^4(\Real^{d_x})$. Since $g_k>0$ (and $\| \phi \|_\infty \le 1$), inequality \eqref{eqP17} yields
$
D(\pi_k^h,\pi_k) \le C_k h
\nn
$, where $C_k = \frac{
	2 \bar C_k h
}{
	\xi_k(g_k)
} < \infty$. \hfill $\QED$

%%%%%%
%
%%%%%%
\section{Sequential ensemble Kalman filter with sequential Euler discretisation} 
\label{apSEnKF}
We outline below the sequential ensemble Kalman filter (SEnKF) with $M$ samples. The algorithm is built around the sequential Euler simulation of the ensemble.

%%%
\begin{enumerate}
\item \textbf{Initialisation}: generate the initial ensemble by drawing $M$ i.i.d. samples $\X_0^\ii$, $i=1, \ldots, M$, from the prior law $\pi_0$. 
\item \textbf{Recursive step:} for $k = 1, \ldots, K$:
	\begin{enumerate}[a)]
	\item {\em Prediction}: for $i=1, \ldots, M$:
		\begin{itemize}
		\item let $X_{n_{k-1}}^\ii := \X_{k-1}^\ii$;
		\item simulate $X_{n_{k-1}+1:n_k-1}^\ii$ using the sequential Euler scheme;
		\item compute the auxiliary states $X_{n_k}^\ii = X_{n_k-1}^\ii + h f_{n_k-1}(X_{n_k-1}^\ii)$,
		\item and set $\hat\X_k^\ii = \hat X_{n_k}^\ii$.
		\end{itemize} 
	\item {\em Sequential Kalman update}:  for $l=0, \ldots, r-1$
		\begin{enumerate}
		\item For $a = j_{l-1}+1, ..., j_l$ and $i=1, \ldots, M$, generate 
		\beq
		\bar\X_{a,k}^\ii = X_{a,n_k-1}^\ii + hf_{a,n_k}( \bar\X_k^\ii[a] ) + \sqrt{h}s_{a,n_k-1}(X_{n_k-1}^\ii) V_{a,n_k}^\ii,
		\nn
		\eeq
		where $j_{-1}=0$, the $V_{a,n_k}^\ii$'s are iid $\mN(0,I_{m_x})$ r.v.'s, $\bar\X_k^\ii[0] = \hat \X_k^\ii$ and, for $a>0$, $\bar\X_k^\ii[a] = \left[ 
			\begin{array}{l}
			\X_{0:j_{l-1},k}\\
			\bar\X_{j_{l-1}+1:a-1,k}\\
			\hat \X_{a:q-1,k}\\
			\end{array}
		\right]$.
		\item Propagate the ensemble members through the observation equation,
		\beq
		Y_{l,k}^\ii = v_l \bar\X_{j_l,k}^\ii, \quad i=1, \ldots, M.
		\nn
		\eeq 
		Compute the mean vector
		$
		y_{l,k}^M = \frac{1}{M}\sum_{i=1}^M Y_{l,k}^\ii, 
		$
		and the covariance matrix
		$
		C_{l,k}^{y,M} = \frac{1}{M-1}\sum_{i=1}^M \left( Y_{l,k}^\ii - y_{l,k}^M \right)\left( Y_{l,k}^\ii - y_{l,k}^M \right)^\top
		$.
		
		\item Let $\tilde\X_{0:j_l,k}^\ii = \left[
			\begin{array}{l}
			\X_{0:j_{l-1},k}^\ii\\
			\bar\X_{j_{l-1}+1:j_l,k}^\ii\\
			\end{array}
		\right]$ and compute the ensemble mean 
		\beq
		\tilde x_{0:j_l,k}^M = \frac{1}{M}\sum_{i=1}^M \tilde\X_{0:j_l,k}^\ii
		\nn
		\eeq
		and the cross-covariance matrix
		\beq
		C_{l,k}^{xy,M} = \frac{1}{M-1}\sum_{i=1}^M \left(
			\tilde\X_{0:j_l,k}^\ii - \tilde x_{0:j_l,k}^M
		\right)\left(
			Y_k^\ii - y_k^M
		\right)^\top.
		\nn
		\eeq
		
		\item Compute the Kalman gain
		$
		G_{l,k}^M = C_{l,k}^{xy,M}\left( C_{l,k}^{y,M} \right)^{-1}.
		$
		
		\item Update the ensemble,
		\beq
		\X_{0:j_l,k}^\ii \leftarrow \tilde\X_{0:j_l,k}^\ii + G_{l,k}^M\left( Y_{l,k} - Y_{l,k}^\ii + \sigma_y U_{l,k}^\ii \right),
		\quad i=1, \ldots, M,
		\nn
		\eeq
		where $\{ U_{l,k}^\ii \}_{i=1}^M$ are iid $\mN(0,I_{d_y})$ r.v.'s. 
		\end{enumerate}
	\end{enumerate}
\end{enumerate}
%%%

%%%%%%
%
%%%%%%
\section{Sequential ensemble Kalman filter with standard Euler discretisation} 
\label{apSEnKF2}
A sequential ensemble Kalman filter with $M$ samples can also be implemented using a standard Euler scheme for the simulation of the ensemble. The algorithm is outlined below.

\begin{enumerate}
\item \textbf{Initialisation}: generate the initial ensemble by drawing $M$ i.i.d. samples $\X_0^\ii$, $i=1, \ldots, M$, from the prior law $\pi_0$. 
\item \textbf{Recursive step:} for $k = 1, \ldots, K$:
	\begin{enumerate}[a)]
	\item {\em Prediction}: for $i=1, \ldots, M$:
		\begin{itemize}
		\item let $X_{n_{k-1}}^\ii := \X_{k-1}^\ii$ and
		\item simulate $X_{n_{k-1}+1:n_k-1}^\ii$ using the standard Euler scheme.
		\end{itemize} 
	\item {\em Sequential Kalman update}:  for $l=0, \ldots, r-1$
		\begin{itemize}
		\item[i.] For $a = j_{l-1}+1, ..., j_l$ and $i=1, \ldots, M$, generate 
		\beq
		\bar\X_{a,k}^\ii = X_{a,n_k-1}^\ii + hf_{a,n_k}( X_{n_k-1}^\ii ) + \sqrt{h}s_{a,n_k-1}(X_{n_k-1}^\ii) V_{a,n_k}^\ii,
		\nn
		\eeq
		where $j_{-1}=0$ and the $V_{a,n_k}^\ii$'s are iid $\mN(0,I_{m_x})$ r.v.'s.
		\item[ii.--v.] Same as in the SEnKF of Appendix \ref{apSEnKF}.
\end{itemize}
	\end{enumerate}
\end{enumerate}
%%%

%%%%%%%%%%%%%%%%%%%%%%%%%%%%%%%%%%%%%%%
%															%
%%%%%%%%%%%%%%%%%%%%%%%%%%%%%%%%%%%%%%%
\bibliographystyle{plain}
\bibliography{bibliografia}

\end{document}